\documentclass[12pt]{article}
\usepackage[dvips]{graphicx}
\usepackage{color}
\usepackage{amsfonts,amssymb}

\begin{document}

\vspace*{1cm}
\begin{center}
{\Large \bf The GZK Puzzle and Fundamental Dynamics}\\

\vspace{4mm}

{\large A. A. Arkhipov\footnote{e-mail: Andrei.Arkhipov@ihep.ru}\\
{\it State Research Center ``Institute for High Energy Physics" \\
 142281 Protvino, Moscow Region, Russia}}\\
\end{center}

\vspace{2mm}
\begin{abstract}
The conjecture that the GZK puzzle might be related with nontrivial
structure of the inelastic defect of total cross sections in
scattering from nuclei has been suggested.
\end{abstract}

\section{Introduction: Ultra High Energy Cosmic Rays\\ and Particle Physics}

From the history of fundamental science everybody knows  that Cosmic
Rays Physics as a part of Astrophysics and Particle Physics
especially based on accelerator studies have many common roots. In
particular, many discoveries early in particle physics have been done
in the study of cosmic rays. It is enough to remind that the
researches in cosmic rays resulted in the discovery of such
elementary particles as the positron $e^+$ in 1932, the muon --
second charged  lepton $\mu^\pm$ in 1937, the charged and neutral
pions $\pi^\pm$, $\pi^0$, the strange particles kaons $K^\pm$, $K_L$,
$K_S$ and $\Lambda$-hyperon in 1947, the antiproton, $\Xi^-$ and
$\Sigma^+$ in 1952-1955.

In the very beginning of the second half of XX century a period of
divergence between Cosmic Rays Physics and Particle Physics, both in
methodology and in the places of interest, has been started. Particle
physicists have taken the path of building the big accelerators and
large detectors. The experiments at the Serpukhov accelerator, the
ISR and the S$p\bar p$S at CERN, the Tevatron collider at FNAL
allowed to learn the hadron interaction properties at high energies.
The accelerator experiments together with theoretical efforts
resulted in the construction of the ``Standard Model''  with clear
understanding and power predictions at least in the electro-weak
sector and with a number of new open questions. It is believed that
new collider experiments such as the LHC project at CERN and others
might help to find the answers to the open questions in Particle
Physics. At the same time it is quite clear that the new measurements
at the accelerator experiments would be of great importance for
Cosmic Rays Physics. This is because high energy cosmic rays  are
usually measured indirectly by investigating the air showers they
produce in the atmosphere of the earth. A correct interpretation of
the air shower measurements with a necessity requires an improved
understanding of the hadron interaction properties, explored by the
accelerator experiments. There is a clear necessity to measure at
accelerators the global characteristics of the high energy hadron
interactions with a high accuracy, in order to accurately interpret
the existing and newly data on the measurements of the highest energy
air showers. Certainly, this request, addressed to the physicists
working on the hadron accelerators, is clearly and strongly
motivated.

The progress in Cosmic Rays Physics during ``accelerator era'' has
been less substantial compared to Particle Physics. Probably the main
reason of that is owing to above mentioned divergence between two
branches in fundamental science. However, the recent results in
cosmic ray studies and new astrophysical observations open a new page
in Particle Physics. One of the most interesting   results is the
detection of cosmic ray particles with energies exceeded
$10^{19}\,$eV.  Ref.~\cite{1} is the first article where the
detection of a cosmic ray with energy $10^{20}\,$ eV has been
published (the Volcano Ranch experiment). At present time the total
number of detected air showers with energy higher than $10^{20}\,$ eV
is about 20 \cite{2}. The existence of cosmic ray particles with
energies above $10^{20}$ eV has been confirmed by all experiments,
regardless of the experimental technique used.\footnote{See, however,
recent article \cite{32}.} Why this result is so interesting for
Particle Physics?

It is well known that soon after the discovery of the Cosmic
Microwave Background Radiation (CMBR) by Penzias and Wilson \cite{3}
almost simultaneously Greisen in the  USA~\cite{4} and
Zatsepin\&Kuzmin~\cite{5} in the USSR predicted that above
$10^{20}\,$ eV the cosmic ray spectrum will steepen abruptly (GZK
effect). The cause of that catastrophic cutoff is the intense
isotropic CMBR which is really a 2.7K thermal blackbody radiation
produced at a very early stage of the Universe evolution and
confirmed by measurements of Roll and Wilkinson  \cite{6}. CMBR
photons with a 2.7K thermal spectrum fill the whole Universe with a
number density of $\sim$ 400 cm$^{-3}$. The physical mechanism of the
GZK cutoff is quite clear, it is based on the interactions of
Ultra-High Energy Cosmic Ray (UHECR) particles with CMBR photons.
Protons, photons, electrons, neutrinos, nuclei etc might be as such
UHECR particles. Mainly the process that cause the energy loss of
UHECR particle, say proton, is the photo-production of pions on CMBR
photons: $p + \gamma^{\rm cmb} \rightarrow p + n\pi$. The threshold
center of mass energy for photoproduction of one pion is $\sqrt
{s_{\rm thr}} = m_p + m_{\pi^0} \simeq$ 1073 MeV. In the Cosmic Rest
Frame (CRF, defined as the frame in which the CMBR is represented as
isotropic photon gas) one can estimate the proton threshold energy
for pion photoproduction
\begin{equation}\label{GZKproton}
E_{p,\,\rm thr}^{\rm CRF}=\frac{s_{\rm
thr}-m_p^2}{2\varepsilon_{\gamma}^{\rm
cmb}}=\frac{m_{\pi}(m_p+m_{\pi}/2)}{\varepsilon_{\gamma}^{\rm
cmb}}\simeq 13.6\times 10^{16}\Big(\frac{\varepsilon_{\gamma}^{\rm
cmb}}{\mbox{eV}}\Big)^{-1}\,\mbox{eV},
\end{equation}
where $E_p^{\rm CRF}$ is the proton energy in the Cosmic Rest Frame.
Taking the average energy $\varepsilon_{\gamma}^{\rm cmb}=6.3\times
10^{-4}\,$ eV,
 the proton threshold energy is $E_{p,\,\rm thr}^{\rm
CRF}\simeq 2.15\times 10^{20}\,$eV.

On the other hand, in the rest frame of a cosmic ray proton (PRF --
projectile rest frame) a substantial fraction of the CMBR photons
will look as $\gamma$--rays with energy above the threshold energy
for pion photoproduction
\begin{equation}\label{GZKphoton}
\varepsilon_{\gamma^{\rm cmb},\,\rm thr}^{\rm PRF}=\frac{s_{\rm
thr}-m_p^2}{2m_p}=m_{\pi}+\frac{m_{\pi}^2}{2m_p}\simeq
145\,\mbox{MeV},
\end{equation}
where $\varepsilon_{\gamma^{\rm cmb}}^{\rm PRF}$ is the CMBR photon
energy in the projectile rest frame. The photoproduction cross
section as a function of projectile photons for stationary protons is
very well measured and studied at accelerator experiments \cite{7}.
There is a detail information that is shown in Fig.~1. At low
energies the cross section exhibits a pronounced resonance associated
with the $\Delta^+$ decaying into $p\pi^0$ mode; here the cross
section exceeds 500 $\mu$b at the peak. The complicated range beyond
the $\Delta^+$--resonance is essentially dominated by the higher mass
resonances associated with multiple pion production $\gamma +
p\rightarrow\Delta^*\rightarrow p+n\pi, n>1$. The whole resonance
range is followed by the long tail with approximately constant cross
section about 100 $\mu$b with a slow increase up to 1 TeV. The
photo-pion production cross section for neutrons is to a good
approximation identical. As seen from Fig.~1 at half-width of the
$\Delta^+$ resonance peak the total cross section $\sigma_{\gamma
p}\simeq$ 300 $\mu$b = 3$\times 10^{-28}\,{\rm cm}^2$. Taking into
account the number density of the CMBR photons $n_{\gamma^{\rm
cmb}}\sim$ 400 cm$^{-3}$, for the mean free path $l$ in the CMBR
photon gas one obtains
\begin{equation}\label{meanfreepath}
l=\frac{1}{n\sigma}\simeq 8.3\times 10^{24}\,{\rm cm}\simeq 2.8\,
{\rm Mpc}.
\end{equation}
The next important parameter is the proton inelasticity $k_{inel}$
defining the fraction of energy that a proton loses in one collision.
At threshold in each collision protons lose about 18\% of their
energy, and this  energy loss fraction increases with increase of
energy. In that way the energy loss length $L=l/k_{inel}$ is
estimated about a few tens Mpc for protons with energy higher than
$10^{20}$ eV.

Apart from photo-pion production, the process of pair production $p +
\gamma^{\rm cmb}\rightarrow p + e^+e^-$ should be considered as well.
While this process has a smaller threshold by a factor
$2m_e/m_{\pi}$, it has a smaller cross section. Thus, the estimated
energy loss length $L$ is about $10^3$ Mpc. Nevertheless, this
process might be important at sub--GZK energies. The detailed
analysis of the energy loss length of protons in interactions with
the CMBR photons is presented in Fig.~2.

For neutrons with $E\sim 10^{20}$ eV, the dominant loss process is
$\beta$--decay $n\rightarrow pe{\bar\nu}_e$. The neutron decay rate
$\Gamma_n=m_n/E\tau_n$, with the laboratory lifetime $\tau_n\simeq
888.6$ sec, gives the neutron energy loss length
\begin{equation}\label{neutronlength}
L_n=\tau_n\frac{E}{m_n}\simeq 0.9\Big(\frac{E}{10^{20}\rm eV}\Big)
{\rm Mpc}.
\end{equation}

Obviously, UHECRs nuclei are expose to the same energy loss processes
as UHECRs protons. So that, the respective threshold for the
photo-pion production reaction, in particular, $A + \gamma^{\rm cmb}
\rightarrow A + \pi^0$, is given by change of $m_p$ with the mass of
the nucleus in Eq.~\ref{GZKproton}
\begin{equation}\label{GZKnucleus}
E_{A,\,\rm thr}^{\rm
CRF}=\frac{m_{\pi}(m_A+m_{\pi}/2)}{\varepsilon_{\gamma}^{\rm
cmb}}\simeq A E_{p,\,\rm thr}^{\rm CRF}.
\end{equation}
From Eq.~\ref{GZKnucleus} it follows that the GZK cutoff energy for
nuclei is shifted to larger values for heavier nuclei. However, it
turns out that the dominant energy loss process for nuclei is
photodisintegration, typically $A + \gamma^{\rm cmb} \rightarrow
(A-1) + N,(A-2)+2N,...$, happening due to giant resonances at about
the same primary energy. In fact, here we have another excellent
example of intersection between nuclear\&particle physics and
high-energy cosmic rays physics. Recent detailed studies reveal that
photodisintegration for nuclei leads to energy loss length of
$\sim$10 Mpc at energy $2\times 10^{20}$ eV, which is comparable with
the energy loss length for nucleons.

Well, to resume the GZK effect physically means that isotropic CMBR
photon gas makes the Universe opaque to UHECRs particles whose energy
is greater than $10^{20}$ eV. In terms of the energy loss length the
GZK cutoff looks like a suppression of the UHECRs flux due to
restriction of the propagation distance to a few tens of Mpc. In that
sense a notion of the GZK sphere arises: simply it is a sphere with
the radius $R_{GZK}\simeq 50$ Mpc within which a source has to locate
to be the origin of the UHECRs particles with energy $\gtrsim
10^{20}$ eV.

As mentioned above the cosmic ray particles with energies exceeded
$10^{20}$ eV have been detected. The data on the UHECRs spectra
measured by Fly's Eye, AGASA, HiRes I, and HiRes II Collaborations
are collected and shown in Fig.~3. extracted from Ref.~\cite{9} (see
references therein). As seen from Fig.~3 the combined UHECR spectrum
does not exhibit the GZK cutoff at all, many events with $E>10^{20}$
eV have been observed. The strongest evidence for trans-GZK events
comes from the AGASA observations. The AGASA group reported the
detection of up to 17 events with energy $\gtrsim 10^{20}$ eV and
claimed that GZK cutoff effect is not observed. Now a non-observation
of the GZK effect is known as the GZK puzzle. Of course, the GZK
puzzle results in uncommonly profound consequences, raising questions
to the nature of the primary UHECRs particles and their sources as
well as the physical mechanisms responsible for endowing cosmic ray
particles with such enormous energies or even to the particle physics
in itself. That's why the GZK puzzle and all around of that are the
targets of wide discussions in the literature at present time. Many
ideas and different models as solutions have been suggested
\cite{10}, however, the true solution of the GZK puzzle is unknown so
far.

Looking at Figure 3, one can see that the UHECR spectrum exhibits a
dip structure at the energy about $10^{20}$ eV, i.e. the spectrum
really has a minimum just at the GZK cutoff energy. Thus the GZK
puzzle (irrespective of the source and accelerating mechanism for
cosmic rays particles) is transformed into the questions: what is the
origin of this minimum, and how could one explain an appearance of
the minimum in the UHECR spectrum in the framework of fundamental
dynamics in particle physics. Here we suggest a conjecture that the
minimum in the UHECR spectrum might be related with nontrivial
structure of the inelastic defect of total cross sections in
scattering from nuclei. This point will be discussed in a more detail
below.

\section{On Total Cross Sections in Scattering from Nuclei}

In the middle of XX century experimental and theoretical studies of
high-energy particle interaction with deuterons have shown that even
in the range of asymptotically high energies the total cross section
in scattering from deuteron cannot be treated as a simple sum of the
proton and neutron total cross sections. Glauber was the first to
explain why a simple addition of the elementary free nucleon cross
sections might be failed. Using the methods of diffraction theory,
the quasiclassical picture for scattering from composite systems and
eikonal approximation for high-energy scattering amplitudes, he found
fifty years ago  \cite{11} that the deuteron total cross section can
be expressed by the formula
\begin{equation}\label{sigmad}
\sigma_d = \sigma_p +\sigma_n - \delta\sigma_d,
\end{equation}
where
\begin{equation}\label{deltasigmaG}
\delta\sigma_d = \delta\sigma_d^G = \frac{\sigma_p\cdot
\sigma_n}{4\pi}<\frac{1}{r^2}>_d.
\end{equation}
Here $\sigma_d, \sigma_p, \sigma_n$ are the total cross sections in
scattering from deuteron, proton and neutron, $<r^{-2}>_d$ is the
average value for the inverse square of the distance between the
nucleons inside a deuteron, $\delta \sigma_d^G$ is the Glauber shadow
correction describing the effect of eclipsing or the screening effect
in the recent terminology. The Glauber shadow correction  has quite a
clear physical interpretation. This correction originates from
elastic rescattering of an incident particle on the nucleons in a
deuteron and corresponds to the configuration when the relative
position of the nucleons in a deuteron is such that one casts its
``shadow" on the other. It is a genuine analog of the effect known
for astronomers; the decrease in luminosity of binary star systems
during eclipses \cite{11}.

Soon after it was  understood that in the range of high energies the
shadow effects may arise due to inelastic interactions of an incident
particle with the nucleons of a deuteron. Therefore, an inelastic
shadow correction had to be added to the Glauber one. A simple
formula for the total (elastic plus inelastic) shadow correction had
been derived by Gribov \cite{12} in the assumption of Pomeron
dominance in the dynamics of elastic and inelastic interactions
\begin{equation}\label{gribov}
\delta\sigma_d^{\Gamma} = 2\int d{\bf q}^2\rho(4{\bf
q}^2)\frac{d\sigma_N}{d{\bf q}^2},
\end{equation}
where $\rho({\bf q}^2)$ is the deuteron (charge) formfactor,
$d\sigma_N/d{\bf q}^2$ is the sum of cross sections of all processes
which take place in interaction of incident hadron with the nucleon
at fixed transfer momentum  ${\bf q}^2$. However, it was observed
later on that the calculations performed by the Gribov formula did
not meet the experimental data: The calculated values of the shadow
correction overestimated the experimental measurements. In that case
the idea, that the Pomeron dominance is not justified at the
accelerator energies, becomes clear.

It was believed sometime that the account of the triple-reggeon
diagrams for six-point amplitude in addition to the triple-pomeron
ones would allow to obtain a good agreement with the experiment. But
careful analysis has shown that discrepancy between theory and
experiment could not be eliminated by taking into account the
triple-reggeon diagrams: In fact, it is needed to modify the dynamics
of the six-point amplitude with more complicated diagrams than the
triple Regge ones \cite{13}. This indicates that up to now there is
no a clear understanding, in the framework of Regge phenomenology,
the shadow corrections in elastic scattering from deuteron.

The main difficulty, which the Regge phenomenology faced with, was
the problem to describe the cross section of the single diffraction
dissociation processes. The latest experimental measurement of $p\bar
p$ single diffraction dissociation at c.m.s. energies $\sqrt{s} =
546$ and $1800\ GeV$, carried out by the CDF group at the Fermilab
Tevatron collider \cite{14},  has shown that the most popular model
of supercritical Pomeron does not describe the existing experimental
data. Recent experimental results from HERA \cite{15} lead us to the
same conclusion. The soft Pomeron phenomenology as currently
developed cannot incorporate the HERA data on structure function
$F_2$ at small $x$ and total $\gamma^{*}p$ cross section from $F_2$
measurements as a function of $W^2$ for different $Q^2$. Such
situation might be qualified as a ``super-crisis'' for the
supercritical Pomeron model. Figure~1 extracted from paper \cite{16}
demonstrates the ``super-crisis'' (see details in Ref.~\cite{18}).

Meanwhile it's quite clear that the theoretical understanding of the
shielding effects in scattering from any composite system is of
fundamental importance, because the structure of shadow corrections
is deeply related to the structure of the composite system itself. At
the same time the structure of the shadow corrections displays new
aspects for the fundamental dynamics.

In the second half of 1970th we have concerned in the study of
dynamics in three particles scattering in some details (see recent
review article \cite{19} and references therein). The
Bethe-Salpeter-type equations reduced to one time have been used as
an implement in our study of a dynamics for the three-body systems.
It turned out that the three-body dynamics, under a consistent
consideration of three-body problem in the framework of local quantum
field theory, with a necessity contained new fundamental forces which
the three-body forces are. The three-body forces in relativistic
quantum theory appear as an inherent connected part of total three
particle interaction which cannot be represented by the sum of pair
interactions. An existence of the three-body forces might be
established even in the perturbation theory expansions. Single-time
formalism in Quantum Field Theory used allows one to give a
constructive definition of the three-body forces beyond the
perturbation theory. On this way it was established that the
fundamental three-body forces are related with specific inelastic
interactions in two-body subsystems of the three-body system, and
they govern the dynamics of special inelastic processes known as
one-particle inclusive reactions.  At the rather common assumptions
we managed to calculate the contribution of the three-body forces to
the deuteron total cross section and to derive the new, extremely
simple and refined formula for defect of total cross section in
scattering from deuteron with clear and transparent physical
interpretation. The obtained structure of the shadow corrections to
the deuteron total cross section has revealed new fundamental scaling
laws \cite{20} in interaction of composite nuclear systems. Here I
would like to concern this point in a more detail.

In our approach the defect of the deuteron total cross section is
represented by the sum of two items
\begin{equation}\label{defect}
\delta\sigma_{d}=\delta\sigma_{d}^{el}+\delta\sigma_{d}^{inel},
\end{equation}
where $\delta\sigma _ {d}^{el}$ is elastic defect, and $\delta\sigma_
{d}^{inel}$ is inelastic one. For the elastic and inelastic defects
one obtains
\begin{equation}\label{partdef}
\delta\sigma_{d}^{el} = 2\,
a_d^{el}(x^2_{el})\,\sigma_{N}^{el},\qquad \delta\sigma_{d}^{inel} =
2\, a_d^{inel}(x^2_{inel})\,\sigma_{N}^{sd},
\end{equation}
where
\[
x^2_{el}=\frac{2B_N^{el}}{R_d^2}=\frac{(R_2^{eff})^2}{R_d^2},\quad
x^2_{inel}=\frac{2B_N^{sd}}{R_d^2}=\frac{(R_3^{eff})^2}{R_d^2}.
\]
As seen from Eq.~\ref{partdef} the elastic defect is proportional to
the total elastic cross section, but the inelastic defect is
proportional to the total single diffraction dissociation cross
section in scattering from nucleon. The proportionality factors
$a_d^{el}$ and $a_d^{inel}$ are called the elastic and inelastic
structure functions of a deuteron correspondingly. Here $R_d$ is the
deuteron radius defined by the deuteron formfactor, and scale
variables $x_{el}$ and $x_{inel}$ are defined through the slope of
forward diffraction cone in elastic scattering $B_N^{el}$ and in
single diffraction dissociation $B_N^{sd}$ which are simply related
to the effective radii of two-body $B_N^{el}=(R_2^{eff})^2/2$ and
three-body $B_N^{sd}=(R_3^{eff})^2/2$ forces. Of course, it is
supposed, that both at elastic and at inelastic interactions with
nucleons of a deuteron, proton and neutron are dynamically
indistinguishable, i.e. appropriate dynamic characteristics for a
proton and neutron are identical
$\sigma_{p}^{el}=\sigma_{n}^{el}=\sigma_{N}^{el}$, $B_p^{el}=
B_n^{el}=B_N^{el}$ etc. Such assumption is quite justified at enough
high energies.

Structure functions have clear and quite a transparent physical
meaning. The function $a_{el}$ is some kind of ``counter'', which
measures out a portion of events related with elastic rescattering of
incident hadron on nucleons of a deuteron among of all the events
during the interaction with a deuteron as whole, and this function
attached to the total probability of elastic interaction of an
incident particle with a separate nucleon in a deuteron. This
function depends on a variable, which is effective radius of elastic
interaction with a nucleon measured with the help of ``scale rule''
with a scale defined by the radius of a deuteron. At each value of
this variable (at a given value of energy) the number of the function
$a_{el}$ determines a weight, which the total cross section of
elastic interaction with a nucleon at given energy enters the defect
of the deuteron total cross section with.

The same physical interpretation with obvious changes in the terms is
transferred on the inelastic structure function. The function
$a_{inel}$ also represents some kind of ``instrument'', but another,
which count out a relative portion of other events among of possible
interactions with a deuteron as a whole related with processes of
inelastic interaction with nucleons inside a deuteron of inclusive
type in the region of diffraction dissociation. The inelastic
structure function depends on another scale variable, which is
effective radius of inelastic interaction with a nucleon measured
with the help of ``scale rule'' with the same scale defined by the
radius of a deuteron. The number of the function $a_{inel}$ at given
value of energy determines a weight,  which the total single
diffraction dissociation cross section on a nucleon at the same
energy enters the defect of the deuteron total cross section with.

Formulas (\ref {defect}) and (\ref {partdef}) may serve as toolkit
for experimental study of structure functions $a_{el}$ and $a_{inel}$
by measurement of the defect for total cross section in scattering
from deuteron with usage of the experimental information about
elastic cross section and total cross section of single diffraction
dissociation on a nucleon. For these purposes, however, it would be
extremely important to have a reliable substantiation of these
formulas. It is remarkable that such theoretical substantiation can
be really obtained.

The formalism, which we have used, allowed us to carry out analytical
calculations completely, if for these purposes to take advantage of
the parameterizations, trustworthy established on experiment, for
differential elastic cross sections and for one-particle inclusive
cross sections in the range of diffraction dissociation
\begin{equation}\label{incspectr}
\frac{d\sigma_N^{el}}{dt}(s,t)=\frac{d\sigma_N^{el}}{dt}(s,0)\exp[B_N^{el}(s)t],
\qquad \frac{2s}{\pi}\frac{d\sigma_N^{sd}}{dtdM_X^2} =
A(s,M_X^2)\exp[b(s,M_X^2)t].
\end{equation}
In this way we managed to get the extremely simple formulas for
structure functions $a_d^{el}$ and $a_d^{inel}$, which look like
\begin{equation}\label{a}
a_d^{el}(x^2)=\frac{x^2}{1+x^2}, \qquad
a_d^{inel}(x^2)=\frac{x^2}{(1+x^2)^{\frac{3}{2}}}.
\end{equation}
It should be especially emphasized once more an important element in
our approach which consists that the inelastic defect in the deuteron
total cross section appears as manifestation of the fundamental
three-body forces, and  at the same the three-body forces determine
the dynamics of one-particle inclusive reactions. Formula relating
the three-body forces amplitude with the one-particle inclusive cross
section has been derived as well; see details in Refs.
\cite{19,20,21} and references therein.

Outcome of functions evaluations (\ref{a}) and analysis of these
functions, however, are worthy of separate discussion. At first,
being returned to the formula (\ref{defect}), we shall remark that
the Glauber formula is followed if in this formula to neglect
inelastic defect and for the elastic structure function to take
approximation $a_d^{el}(x^2) \simeq x^2 $ justified at $x^2 << 1$,
and to take into account that $ \sigma_{N}^{el} \simeq
\sigma_{N}^{tot\, 2}/16\pi B_N^{el}$. Secondly, it is necessary to
pay attention that the structure functions $a_d^{el}$ and
$a_d^{inel}$ have quite different behavior: $a_d^{el}(x^2) $ is the
monotonic (increasing) function when argument vary on a semi-infinite
interval $0\leq x^2<\infty$, and the range of its values is limited
to an interval $0 \leq a_d^{el} < 1$, while the function $a_d^{inel}$
at first increases, reaches a maximum at $x^2=2$, and then decreases,
disappearing at infinity, thus the range of its values is an interval
$0 \leq a_d^{inel} \leq 2/3\sqrt {3}$. Certainly, that such
distinction in behavior of structure functions $a_d^{el} $ and
$a_d^{inel} $ results in far-reaching physical corollaries. For
example, at superhigh energies corresponding $x^2\rightarrow\infty$,
we discover the effect of weakening the inelastic screening i.e. the
inelastic defect disappears (taking into account that
$\sigma_N^{sd}<{\rm Const, s\rightarrow\infty}$), the elastic defect
tends to doubled value of the nucleon total elastic cross section,
and the deuteron total cross section comes nearer to doubled value of
the nucleon total absorption cross section. Therefore, at superhigh
energies the $A$--dependence of the total cross sections in
scattering from nuclei should be recovered with that only by odds
that the fundamental value, standing at $A$, is not the nucleon total
cross section but the nucleon total absorption one. This means that
the total absorption (inelastic) cross section manifests itself as a
fundamental dynamical quantity for the constituents in a composite
system at superhigh energies.

Of course, without any doubt, matching of the obtained theoretical
outcomes with available experimental data on total cross sections in
scattering of protons and antiprotons from deuterons represented for
us the special interest. In figures 5 and 6 the preliminary results
of such matching are shown. The curves in these figures correspond to
the total cross sections in scattering of protons and antiprotons
from deuterons calculated by the formulas (\ref{sigmad},
\ref{defect}, \ref{partdef}, \ref{a}). There the global descriptions
of $pp$ and $\bar pp$ total cross sections (see Figs.~7,8) as well as
of total single diffraction dissociation cross section in view of the
latest experimental data obtained by CDF Collaboration at FNAL \cite
{14}, made by us earlier \cite {18}, have been used.

Besides in the given occasion it should be necessary still to
emphasize, that the matching with experimental data on total cross
sections in scattering of protons and antiprotons from deuterons was
carried out, as it were, in two stages. At the first stage the
theoretical calculations were compared to experimental data on the
antiproton-deuteron total cross section on the supposition, that
$R_d^2 $ is the single free parameter, which value should be
determined from the fit to experimental data. As a result of a
statistical analysis the following value for the $R_d^2$ was
obtained: $R_d^2=66.61 \pm 1.16 \, {\rm GeV}^{-2}$. Here pertinently
to pay attention to the following circumstance. The last experimental
measurements of the deuteron matter radius testify $r_{d, m} =1.963
(4) \, \rm fm$ \cite {22} whence follows that $r^2_{d, m} = 3.853
\,{\rm fm}^2 = 98.96 \,{\rm GeV}^{-2}$. The obtained value for $R_d^2
$ satisfies equation $R_d^2 = 2/3 \, r^2_{d, m}$. For entirety, the
outcomes of theoretical calculations are represented in Fig. 7 up to
energies of Tevatron at FNAL. At the second stage the experimental
data on the proton-deuteron total cross section were compared to
theoretical calculations, in which the value of $R_d^2$ was fixed on
the numerical value, which was obtained at the first stage from the
analysis of the data on $ \bar pd$ total cross section. In other
words, the curve in Fig. 8 corresponds to theoretical calculations
made with the help of the formulas (\ref{sigmad}, \ref{defect},
\ref{partdef}, \ref{a}), in which there was no free parameter. In
this figure the outcomes of theoretical calculations are also
represented up to energies of Tevatron at FNAL. As is seen, the
figures 7 and 8 testify to the excellent agreement between the theory
and experiment. In addition Figures 9 and 10 demonstrate our global
description of the proton-proton total cross section from the most
low energies up to energies reachable in cosmic rays.

In Figure 11 the outcomes of the theoretical calculations of elastic
and inelastic defects of total cross section in scattering of
(anti)protons from deuterons in energy range $\sqrt{s}\sim 10\div
2000$ GeV have been depicted. It follows from these calculations that
the value of elastic defect makes about 10\% from the value of
nucleon--nucleon total cross section, and the value of inelastic
defect makes about 10\% from the value of elastic defect i.e.
approximately 1\% from the value of nucleon--nucleon  total cross
section. Figuratively expressing, it would be possible to tell that
if the elastic defect represents a fine structure of total cross
section in scattering from deuteron then the inelastic defect should
be referred to a hyperfine one.

In our approach the inelastic defect is related to manifestation of
fundamental three-body forces, therefore in this sense the three-body
forces play a role of ``fine tuning'' in the dynamics of the
relativistic three-particle system. It is necessary to render homage
to the physicists-experimenters creating setups with the accuracy of
measurements permitting to discriminate inelastic defects in total
cross sections of particles scattering at high energies. In this
connection the further experimental precise  measurements of the
hadron--deuteron total cross sections at high energies seem to be
extremely important

As it was already mentioned above the maximum value of inelastic
defect is achieved at $(x_d^{inel})^2=2$ ($(x_d^{inel})^2\equiv
R_3^2/R_d^2$). Or else, the value of energy corresponding to maximum
value of the inelastic defect is defined from the equation
$R_3^2(s_{\rm max}) =2R_d^2$. The calculations made in view of our
analysis of existing experimental data give $\sqrt{s_{\rm
max}}=9.01\times10^8\,\mbox{GeV}=9.01\times 10^{17}\mbox{eV}$.
Reevaluating c.m. energy $\sqrt{s_{\rm max}}$ to the lab. system one
obtains $E_{\rm max}^{\rm lab}\simeq s_{\rm max}/(2m_p)=4.8\times
10^{25}{\rm eV }$. It is obvious, that such values of energies are
not accessible on current and design accelerators. However it would
be extremely interesting to look for manifestations of the given
effect in phenomena related with extremely high energy cosmic rays.

\section{Discussion}

The theoretically calculated inelastic defect in the region of a
maximum is shown in Figure 12. As it follows from Eq.~(\ref{sigmad})
a maximum of the inelastic defect corresponds to a minimum of the
total cross section. We have plotted in Figure 13 the
(anti)proton--deuteron total cross section scaled by the factor
$\ln^{3/2}(\sqrt{s/s_0})$ in the region of a maximum of the inelastic
defect. The scale factor is selected {\it ad arbitrium} for a goal of
illustration only to discern a minimum in the total cross section.

Let's remark, however, that the value $s _ {\rm max} $ has clear
physical meaning, it separates two ranges on energy: the range of
energies $s < s_{\rm max}$, at which effective radius of three-body
forces does not exceed size of a deuteron or more exactly $R_3^2(s)/2
< R_d^2$, and the range of energies $s > s_{\rm max}$, at which
effective radius of three-body forces becomes more than size of a
deuteron $R_3^2(s)/2 > R_d^2 $. The existence of boundary $s_{\rm
max}$, since which there is a suppression of inelastic defect, seems
to be the extremely important characteristic of fundamental dynamics.

Here  we would like to make a conjecture that the observed structures
in the cosmic rays spectra, in particular a minimum in the UHECR
spectrum, might be related with the existence of such boundary.

From the Glauber formula (\ref {deltasigmaG}) it follows, that with
decrease of inter-nucleon distance in a deuteron the value of elastic
defect grows. But the configurations with small inter-nucleon
distances in a deuteron are most favorable for a manifestation of
purely three-particle interaction. When effective interaction radius
of an incident hadron with a nucleon becomes comparable with
inter-nucleon distance, the pattern of elastic rescattering on
nucleons of a deuteron ceases to be adequate to complete pattern of
interaction with a deuteron. In this case it is also necessary  to
take into account purely three-particle forces. It is obvious, that
in a deuteron the configurations are dynamically probable, when the
nucleons are close from each other, but the Glauber theory does not
allow to take into account such configurations. Account of such
configurations demands a more detailed study of the dynamics of
processes of scattering from a deuteron. The technique of the dynamic
equations in a quantum field theory, which we have used, just allows
to carry out such detailed investigations. Once again it should be
emphasized, that the important role in our researches was assigned to
conceptual notion of fundamental three-body forces which with
necessity arise by consistent consideration of the dynamics of three
particles system within the framework of a relativistic quantum
theory. The relation of fundamental three-body forces with dynamics
of one-particle inclusive reactions represents the important outcome
obtained, as it were, by the way.  This outcome especially is
important, that can form the basis both for elaboration of methods of
analytical calculations and for a different sort of phenomenological
analysis.

The comparison of the theory with experimental data on
(anti)proton--deuteron total cross sections made shows, that for the
description of particles scattering from a deuteron at high energies
it is enough to take into account only nucleon degrees of freedom in
a deuteron. The weakly bounded two-nucleon system the deuteron looks
so, that the clusterization of quarks in nucleons is not broken even
then, when the nucleons approach closely to each other. Nucleons,
being close from each other in a deuteron, do not lose of the
individuality and consequently there is no necessity to introduce the
six-quark configurations depersonalized in a deuteron. The structure
derived for the defect of total cross section in scattering from a
deuteron corresponds to such pattern.

We managed to show, that the general formalism of quantum field
theory admits a possibility of representation of dynamics of a
particle scattering from composite system through the fundamental
dynamics of a particle scattering from isolated constituents and
structure of the composite system itself. Though the dynamics of a
particle scattering from two-particle bound system a deuteron was
considered in details, the general formalism used admits a natural
generalization and extension to more complex multiparticle compound
nuclear systems. Certainly, the complexity of consideration, at that,
substantially increases.

Really, to consider the problem of scattering from nucleus consisting
of $A$ nucleons we have to solve many-body problem for
($A+1$)--particle system. However, instead of solving this very
complicated problem one could use a powerful reduction method. For
this goal let's consider a nucleus consisting of $A$ nucleons as a
bound system of one nucleon and nucleus consisting of ($A-1$)
nucleons. In that case the problem of scattering from a nucleus with
$A$ nucleons is reduced to the problem of scattering from a two-body
bound system which has  been previously solved. Of course such
supposition is not unique and should be considered as a some sort of
simplification. In the other way one could suppose that a nucleus
consisting of $A$ nucleons may be represented as a two-body bound
system of a nucleus with $A_1$ nucleons and other nucleus with $A_2$
nucleons so that $A_1+A_2=A$. By this way the problem of scattering
from a nucleus with $A$ nucleons is also reduced to the problem of
scattering from a ``deuteron'' previously solved. Anyway continuing a
reduction in both cases we will come at a final stage to the
expression for total cross section in scattering from a nucleus in
terms of fundamental dynamics in scattering from a nucleon and the
structure of a nucleus.

Formula (\ref{sigmad}) for total cross section in a case of
scattering from any nucleus with $A$ nucleons can be rewritten in the
form
\begin{equation}\label{sigmanucl}
\sigma_A = A \sigma_N  - \delta\sigma_A,
\end{equation}
where $\sigma_N$ is the total cross sections in scattering from
nucleon. The defect $\delta\sigma_A$, in general, also contains two
parts as in Eq.~(\ref{defect})
\begin{equation}\label{defectnucl}
\delta\sigma_{A}=\delta\sigma_{A}^{el}+\delta\sigma_{A}^{inel},
\end{equation}
where $\delta\sigma _ {A}^{el}$ is elastic defect, and $\delta\sigma_
{A}^{inel}$ is inelastic one. From general point of view, as
presented above, for the elastic and inelastic defects one can write
\begin{equation}\label{partdefnucl}
\delta\sigma_{A}^{el} = a_A^{el}(X^2_{el})\,\sigma_{N}^{el},\qquad
\delta\sigma_{A}^{inel} = a_A^{inel}(X^2_{inel})\,\sigma_{N}^{sd},
\end{equation}
where
\[
X^2_{el}=\frac{2B_N^{el}}{R_A^2}=\frac{(R_2^{eff})^2}{R_A^2},\quad
X^2_{inel}=\frac{2B_N^{sd}}{R_A^2}=\frac{(R_3^{eff})^2}{R_A^2}.
\]
The functions $a_A^{el}$ and $a_A^{inel}$ are called the elastic and
inelastic structure functions of a nucleus. Here we have included the
combinatorial factors in the definition of the structure functions.
$R_A$ is the nucleus radius defined by the nucleus formfactor. The
scaled variables $X_{el}$ and $X_{inel}$ are defined as $x_{el}$ and
$x_{inel}$ above but with another scale factor $R_A$ which is the
radius of a nucleus as it should be. It is obviously that formulas
(\ref {defectnucl}) and (\ref {partdefnucl}) may serve as a base to
experimentally  study the structure functions $a_A^{el}$ and
$a_A^{inel}$ by measuring the defect of total cross section in
scattering from nuclei with using the experimental information about
elastic cross section and total cross section of single diffraction
dissociation in scattering from a nucleon. To calculate the structure
functions $a_A^{el}$ and $a_A^{inel}$ is a task of any theory or
theoretical model. As mentioned above the calculation of  the
structure functions $a_A^{el}$ and $a_A^{inel}$ in quantum field
theory is a very complicated problem. However, it would reasonably to
use our experience acquired in solution of this problem for deuteron
case.

Really, it seems to a good approximation, one could use the following
expressions for the structure functions $a_A^{el}$ and $a_A^{inel}$
of any nucleus
\begin{equation}\label{a_A}
a_A^{el}(X^2)=A\frac{X^2}{1+X^2}, \qquad
a_A^{inel}(X^2)=A\frac{X^2}{(1+X^2)^{\frac{3}{2}}}.
\end{equation}
Here we have applied the identity
\begin{equation}\label{combinatorics}
\sum_{k=2}^{A}(-1)^k k {A\choose k}=A,
\end{equation}
and supposed that any many-fold rescattering in a nucleus feels one
and the same structure function like for two-fold rescattering. It is
clear that this supposition is a strong enough simplification,
however, it might  be precise one  at ultra-high energies. At any
rate, it would be very desired to test such simple pattern in
ultra-high energy cosmic rays.

For the effective radius of three-nucleon forces obtained in our
previous investigations one can write the following analytical
expression
\begin{equation}\label{3Nradius}
R_3^2(s)=\Big[5.2667 + 0.4137\ln^2(s/s_0)^{1/2}\Big]\, {\rm
GeV}^{-2},\quad ({s_0})^{1/2}=20.74 \,{\rm GeV}.
\end{equation}
Then the equation $R_3^2(s_{\rm max})=2R_A^2$, defining a value of
energy at a maximum of the inelastic defect, has an obvious solution
\begin{equation}\label{smax_A}
({s_{\rm
max}})^{1/2}=({s_0})^{1/2}\exp\Bigg[\frac{2R_A^2-5.2667}{0.4137}\Bigg]^{1/2}.
\end{equation}
Further, if we put as $R_A=r_0A^{1/3}=R_d(A/2)^{1/3}$, then
Eq.~(\ref{smax_A}) can be rewritten in the form
\begin{equation}\label{smax_An}
({s_{\rm
max}})^{1/2}=20.74\exp\Bigg[\frac{133.22(A/2)^{2/3}-5.2667}{0.4137}\Bigg]^{1/2}\,{\rm
GeV},
\end{equation}
where the value for the deuteron radius $R_d^2=66.61\,\rm GeV^{-2}$
mentioned above has been used. From Eq.~(\ref{smax_An}) one obtains
\[
({s_{\rm max}})_{A=2}^{1/2}=9.01\times 10^8\,{\rm GeV},\, ({s_{\rm
max}})_{A=3}^{1/2}=1.27\times 10^{10}\,{\rm GeV},\, ({s_{\rm
max}})_{A=4}^{1/2}=1.03\times 10^{11}\,{\rm GeV},...
\]
Here, it is interesting to note that the GZK cutoff value $\sim
10^{11}$ GeV appears for c.m. energy corresponding to maximum of the
inelastic defect in a case of scattering from Helium nucleus.

The $A$-dependence in position of a maximum of the inelastic defect
in scattering from nuclei may have a direct link to the question on
chemical composition of the cosmic rays. An information about
chemical composition of the cosmic rays can be elicited from detailed
study of air showers development. It is quite clear, for instance,
that air showers produced by heavy nuclei start the development in
the earth atmosphere earlier compared to protons as primaries. The
cosmic rays composition is often investigated by fitting the energy
dependence of the depth into the atmosphere of maximum $X_{\rm max}$
of the UHECRs-generated air showers. In fact, $X_{\rm max}$ is the
atmospheric depth at which the number of particles in a shower
reaches its maximum. This quantity strongly depends on the primary
energy and composition, that's why $X_{\rm max}$ if often considered
as the most useful observable of the air showers. At ground array
detectors $X_{\rm max}$ is mainly provided by measuring the muon
content or more exactly the ratio of electrons to muons in air
shower. In other case, optical fluorescent detectors allow to
directly observe air shower development. Just to say qualitatively,
it should be mentioned that for a given primary energy a heavier
nucleus creates air shower with a higher muon content and $<X_{\rm
max}>$ is higher up in the atmosphere compared to those for a
proton-generated air shower. The higher muon content of air shower
produced by heavy nucleus can be understood by the fact that it is
relatively easier for charge pions to decay to muons before
interacting with the medium when the shower develops higher up in the
less dense atmosphere. Besides, a less energetic pions generated from
heavy nucleus have a higher decay probability, therefore the muon
fraction is higher in air showers produced by heavy nuclei as well by
this reason. This is clearly demonstrated in Fig.~14 where the
results of the Kascade air shower experiment have been fitted, using
the QGSJET model generator, with a composition dominated by Helium
nuclei and smaller contributions of proton, ${}^{16}$O and
${}^{56}$Fe.

As can be seen from Figure 15, the iron fraction gradually decreases
when changing the energy from $10^{17}$ to $10^{20}$ eV, but the
fraction of lighter nuclei increases in the same interval of energies
i.e. there is trend from heavy toward lighter composition in the
measurements of  $<X_{\rm max}>$. It seems, the recent studies
indicate that at the highest energies $\gtrsim 10^{19}$ eV there is a
significant fraction of nuclei with charge greater than unity, and
less than 50\% of the primary cosmic rays can be photons. In another
words, the existing experimental data suggest that UHECRs are
predominantly protons or light nuclei as for cosmic rays at much
lower energies. However, due to poor statistics and large
fluctuations from shower to shower the definite conclusions on the
composition of the UHECRs have to await data from next generation
experiments.

Here, we would like to emphasize that the simulation of shower
development depends on the event generator used containing some model
of hadronic interaction which results in further complication of data
interpretation. The major uncertainties in air shower simulation stem
from the hadronic interaction models which are usually represented by
empirical parameterizations, and therefore almost all hadronic models
are purely phenomenological. This is because one cannot calculate
soft hadronic interaction cross sections or hadronic multiparticle
production within QCD from first principles. The second major source
of uncertainty is the large extrapolation (over 6 orders of magnitude
in energy) from accelerator experimental data to the UHECRs ones. In
this place the reliable model and the precise accelerator data on
fundamental, for example hadron-proton, total cross sections as well
as on total cross sections in scattering from nuclei are needed to
constrain uncertainties in the interpretation of cosmic rays data to
accurately determine the energy spectrum and the composition of the
UHECRs. Additionally, theoretical understanding and description of
the diffractive dissociation processes are of special importance for
nuclei interactions and consequently for air shower development too.
The most popular model which has been used to simulate interactions
of nucleons and nuclei is based on Regge phenomenology with the
super-critical Pomeron exchange. However, as mentioned above, this
model faced with a serious difficulties in description of single
diffractive dissociation in $p\bar p$ collisions. Moreover, the
super-critical Pomeron model breaks  the fundamental principles of
relativistic quantum theory such as unitarity, and this fact is often
overlooked. But only this pathology of the super-critical Pomeron
model is enough to reject the model from consideration. Recent
accurate and complete analysis of experimental data on hadron total
cross sections rejects this model from statistical point of view.
Another, and sometime neglected, source of uncertainties is
uncertainty provided by the measurements of $p\bar p$ total cross
sections performed at Tevatron. Really, the CDF Collaboration
\cite{25} obtained $\sigma_{p\bar p}^{tot}=81.83\pm 2.29$ mb which is
considerably greater than those reported by E710 ($72.81\pm 3.1$ mb)
\cite{26} and E811 ($71.71\pm 2.02$ mb) \cite{27}. Such difference in
the measurements permits of a wide range of different extrapolations.
Of course, the arising uncertainty directly transfers to predictions
for air showers. Namely, the main source of uncertainty of air shower
predictions comes from differences in modelling hadronic interactions
which cannot be eliminated by existing accelerator data.  That is why
an accurate measurement of the $pp$ total cross section at LHC is of
great importance since it would allow to discriminate the different
extrapolations and to make a selection among currently used models.
The study  of $pA$ (proton-nucleus) collisions for the light nuclei
at LHC is also of greatest interest, of course, not only for
fundamental particle physics but for air shower physics as well.

At the same time it's quite clear, that Quantum Field Theory provides
a sound theoretical basis with a definite guidelines how the
fundamental interactions evolve with energy. In this respect the
discussed here global description of the hadron total cross sections
performed in the framework of general structures of local quantum
field theory keeps a preferable place. Thus an incorporation of the
global pattern of hadronic interactions in generally used generators
of events would be extremely desired.

The next widely discussed subject is the question of origin of cosmic
rays. Although cosmic ray particles were discovered almost one
hundred years ago since the first announcement of their observation
in 1912 \cite{28}, the problem of origin of cosmic rays especially of
UHECRs particles has no solution so far. The total cosmic rays
spectrum is shown in Fig.~16. The commonly accepted point of view is
that at energies below 1 GeV the cosmic rays spectrum is dominated by
particles coming from the Sun because the intensities at such
energies are correlated with the Solar activity. At higher energies
between 1 GeV and up to the knee region (see Fig.~16) there are
several arguments including energetics that an origin of the cosmic
rays is outside the Solar system but confines yet to the Galaxy. At
still higher energies between the knee and the ankle, and finally,
beyond $10^{19}$ eV the situation becomes unclear, although the
UHECRs are generally expected to have an extragalactic origin due to
apparent isotropy, and the ankle is sometimes interpreted as a cross
over from Galactic to extragalactic component. At any rate, it is
generally believed that the bulk of the cosmic rays observed at the
Earth is of extra-Solar origin.

Here we would like to suggest quite another new idea that the bulk of
the cosmic rays observed at the Earth is of solely Solar origin. In
particular, the UHECRs particles coming to the atmosphere of the
Earth might be produced by reaction
\begin{equation}\label{uhecrprod}
X_{\otimes} + \bigodot\rightarrow P_{\odot} + \bigodot,
\end{equation}
where we have used the notations: $X_{\otimes}$ for Galactic or
extra-Galactic UHECRs particle, $\bigodot$ for the Sun, $P_{\odot}$
for UHECRs particle coming to the atmosphere of the Earth from the
Sun. Due to reaction (\ref{uhecrprod}) almost all energy of
$X_{\otimes}$ particle is transferred to $P_{\odot}$ particle. The
idea is based on the fact that the cross section of reaction
(\ref{uhecrprod}) is in 4 orders greater than the cross section of
direct interaction of $X_{\otimes}$ particle with the Earth. Of
course, we did not concern of what is the source of the UHECRs in the
Universe, one only claims that UHECRs particles coming to the
atmosphere of the Earth are produced on the Sun.

In this respect we would like to remind an old idea suggested in
\cite{30} to consider the Virgo cluster as a source of the UHECRs.
According to this idea the UHECRs particles, generated in M87 galaxy
in Virgo cluster, diffuse from the center of Virgo in a postulated
extragalactic field with the energy dependent diffusion coefficient,
and they are focusing to the Sun by galactic magnetic field. It is
remarkable that there is no GZK cutoff, and there is no large
anisotropy in the model. Recently a new revival of this very
interesting idea has been proposed. Introducing a simple Galactic
wind in analogy to Solar wind it has been shown \cite{31} that
back-tracing the orbits of the highest energy cosmic rays events
suggests that they may all come from the Virgo cluster, probably from
the active radio galaxy M87. Figure 17 shows the directions of the
cosmic rays events at that point when they leave the halo of our
Galaxy in polar projection. In Fig.~17 the direction to the active
Galaxy M87 (Virgo A), which is the dominant radio Galaxy in the Virgo
cluster, is pointed out as well for reference. The two highest energy
events are shown in Fig.~17 twice: in assuming (i) that they are
protons, and (ii) that they are Helium nuclei (filled black symbols).
The shaded band in Fig.~17 corresponds to the supergalactic plane. A
remarkable observation made in the model calculations is that the
directions of all tracks point North \cite{31}. With exception of two
events with highest energy, all other 11 events can be traced to
within less $20^{\,\circ}$ from Virgo A. Considering  the uncertainty
of the actual magnetic field distribution, it was found that all
events are consistent with arising originally from Virgo A. Besides,
if the two highest energy events are really Helium nuclei, then all
13 events point within 20 degrees of Virgo A. Of course, it is very
interesting that the simple model for a Galactic wind rather similar
to the Solar wind may allow particle orbits at $10^{20}$ eV to be
bent sufficiently to allow ``trans-GZK'' particles to arrive to the
Sun from Virgo from different directions in agreement with the
apparent isotropy in arrival directions. If the model assumptions
might be confirmed then all powerful radio-galaxies might be
considered as sources of the UHECRs. In that case the fantastic idea
arises to use the powerful radio-galaxies as gigantic accelerators to
set up particle interaction  experiments in the sky \cite{31}.  The
Sun may be used as a target, the atmosphere of the Earth -- as a
calorimeter to detect the highest energy events.

\section{Conclusion}

We did not intend in this article to present a full understanding the
cosmic rays observations. Really we have only concerned very special
but at the same time quite intriguing observation related to absence
(GZK puzzle) of the predicted catastrophic cutoff (GZK effect) of the
UHECRs spectrum at energy value about $10^{20}$ eV. Without any
doubt, an observation of a significant flux of UHECRs particles with
energy above the expected GZK cutoff value is of great interest, and
many attempts have been undertaken to explain the existence of such
particles. As mentioned above an explanation of these particles
requires the existence of extremely powerful sources within so called
GZK sphere with the radius about a few tens Mpc. There is evidence
that such powerful radio-galaxies may really exist although this fact
should be clearly confirmed by future experiments with the higher
statistics.

It should be fair to note that there is a controversial  point of
view which means that there are no particles with energy above the
GZK cutoff value, but the present results of AGASA and others, where
such particles have been observed, are artefact of a combination of
incorrect energy calibration, larger than predicted fluctuations in
shower development, non gaussian tail in measurements etc. We did not
touch this point of view  at all. But here it should be pointed out
the recent article of the HiRes Collaboration \cite{32} where the
HiRes measurement of the flux of ultrahigh energy cosmic rays with
fluorescence technique shows a suppression at an energy of $6\times
10^{19}$eV, exactly the expected cutoff energy. The statistical
significance of the break in the spectrum identified with the GZK
cutoff is $\sim 5\sigma$. The measured energy of the cutoff is
$(5.6\pm 0.7\pm 0.9)\times 10^{19}$eV, where the first uncertainty is
statistical and the second is systematic. At the same time Teshima
(for AGASA Collaboration) \cite{33} presented at the International
Conference on High Energy Physics ICHEP2006 (Moscow, Russia, July
26--August 2, 2006) a new (preliminary) AGASA reanalysis with recent
CORSIKA M.C. in which the number of events above $10^{20}$eV was
reduced from 11 to 5$\sim$6, and the flux difference between AGASA
and HiRes became less significant. Nevertheless the main conclusion
in summary of the talk given by Teshima has been unchanged: ``Super
GZK particles exist''. This means that the GZK puzzle exists as well.
In fact, the new (preliminary) AGASA data extend beyond the GZK
cutoff energy with no apparent suppression and probably without dip
and bump. However, it should be noted that the combined AGASA\&HiRes
data presented in Fig. 18 (Fig. 3 from Ref. \cite{32}) clearly show
the dip structure at the GZK cutoff energy as mentioned in the
Introduction.

We also did not concern any exotic models to explain the existence of
``trans-GZK'' particles. Among them there are the so called
``Z-burst'' scenario, Top-Down models, and even super-exotic
explanation due to violation of Lorenz invariance; see e.g. excellent
review article \cite{34} and references therein. Our conjecture is
the attempt to find the solution of the GZK puzzle in the fundamental
dynamics of scattering from nuclei. Certainly, it needs to do much
work else to convert the conjecture into strong statement. At the
same time it is expected that next generation experiments in
Astrophysics, especially in Cosmic Rays Physics, will be able to
yield significant new information about fundamental processes in
Particle Physics. Here the measurement of the UHECRs spectrum beyond
the GZK cutoff is of great importance because such measurements being
a grand element of Cosmic Rays Physics has a very deep relation with
Particle Physics.

It seems the upcoming researches of particle interactions (in the sky
and on the Earth) promise to represent the near future as an exciting
epoch in science when the three branches of science Cosmology,
Astrophysics and Particle Physics are very probably to be combined
into unique common fundamental science such as
CosmoAstroPartPhysology or CASP-Physics. Certainly, we believe in it.

%\newpage
\vspace*{10cm}
\begin{figure}[h]
\vspace*{3cm}
\begin{center}
\includegraphics[width=\textwidth]{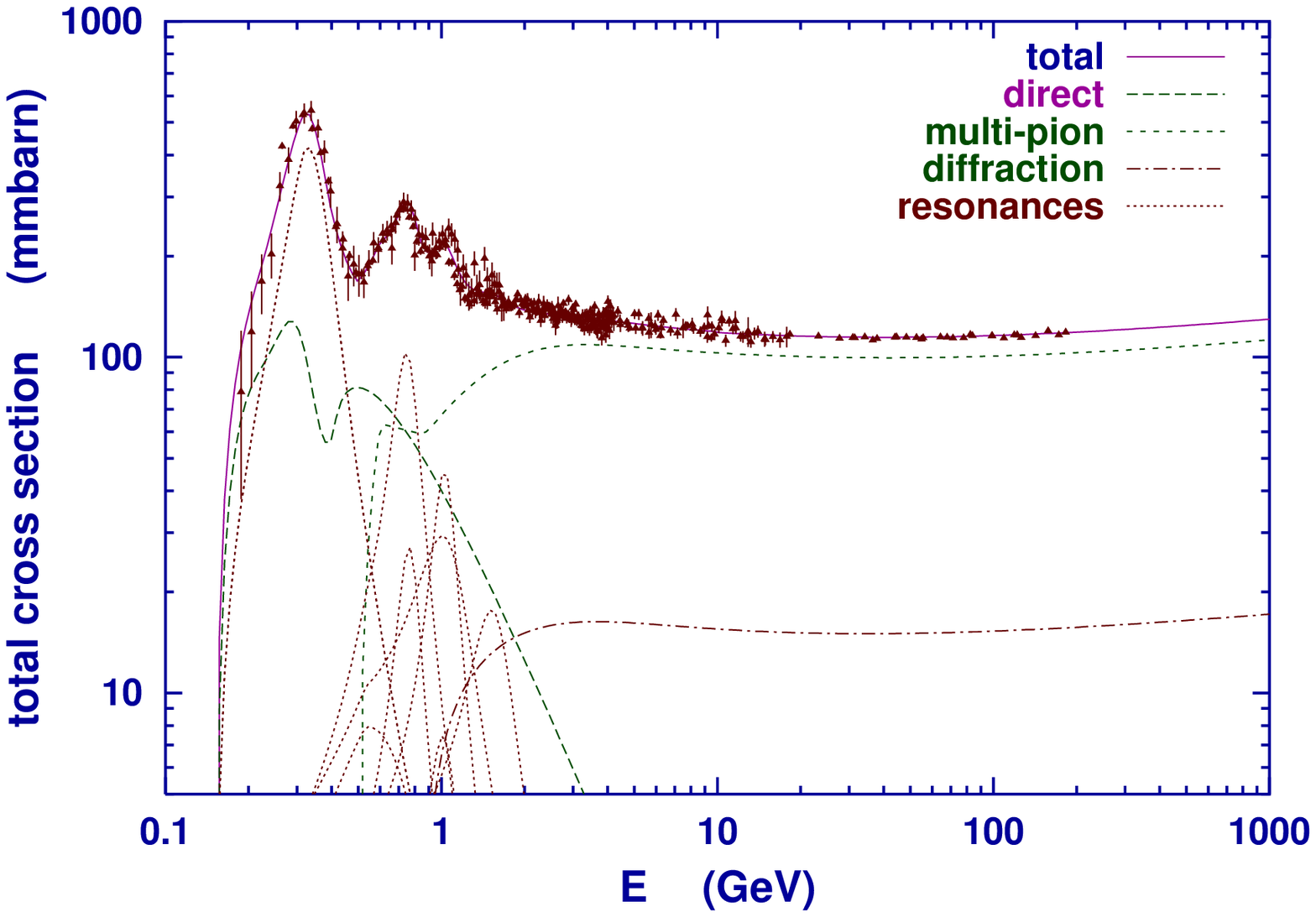}\label{fig1}
\end{center}
\caption{The total photo-pion production cross section for protons as
a function of the photon energy in the proton rest frame (from
Ref.~\cite{8}).}
\end{figure}

%\newpage
\vspace*{3cm}
\begin{figure}[htb]
\begin{center}
\includegraphics[width=\textwidth]{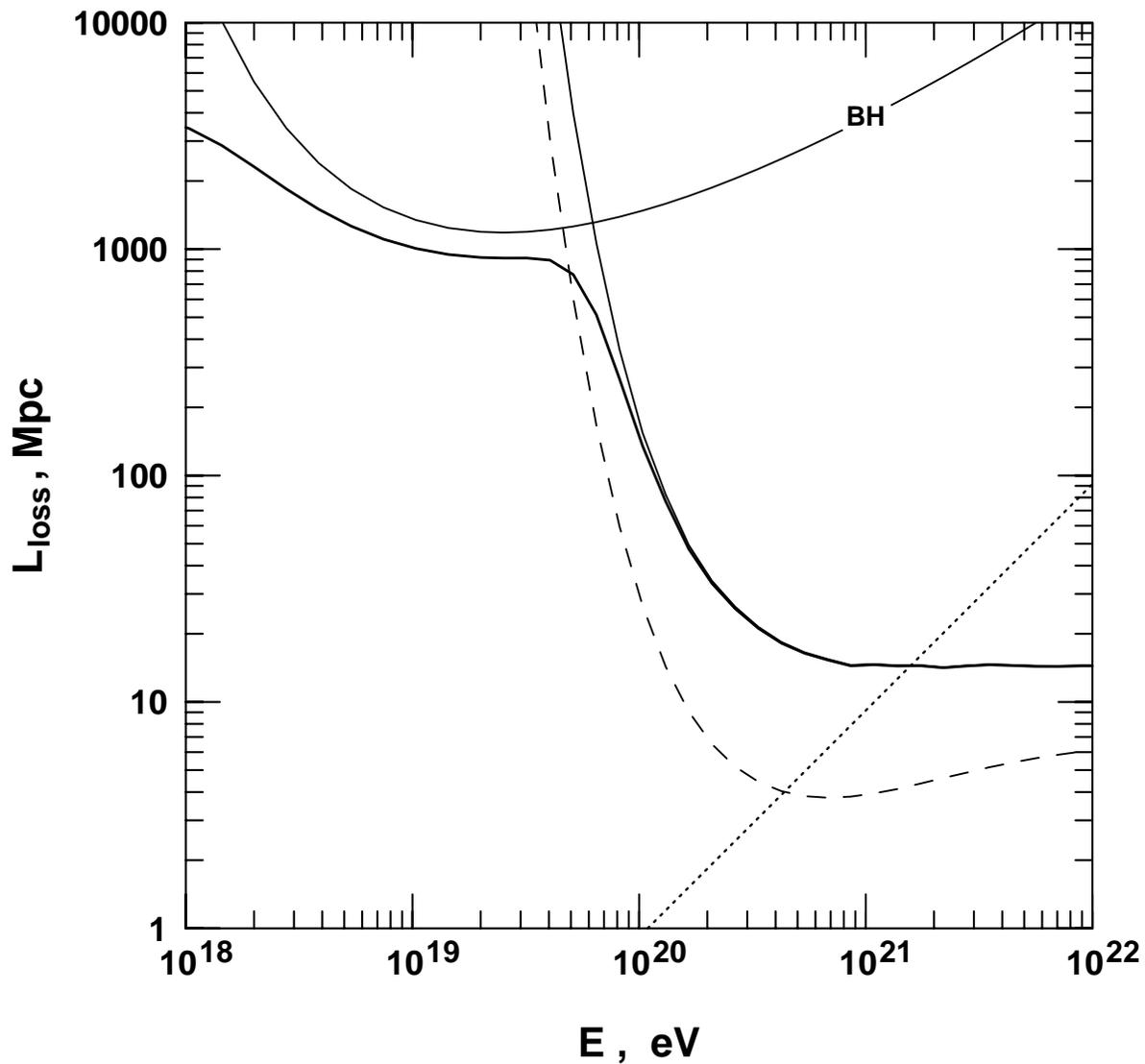}\label{fig2}
\end{center}
\caption{Energy loss length of protons in interactions with the CMBR
photons (thick solid line). The dashed line shows the proton
interaction length. The contribution of the pair production is shown
with a thin line. The dotted line shows the neutron decay length
(from Ref.~\cite{8}).}
\end{figure}

%\newpage
\begin{figure}[htb]
\begin{center}
\includegraphics[width=\textwidth]{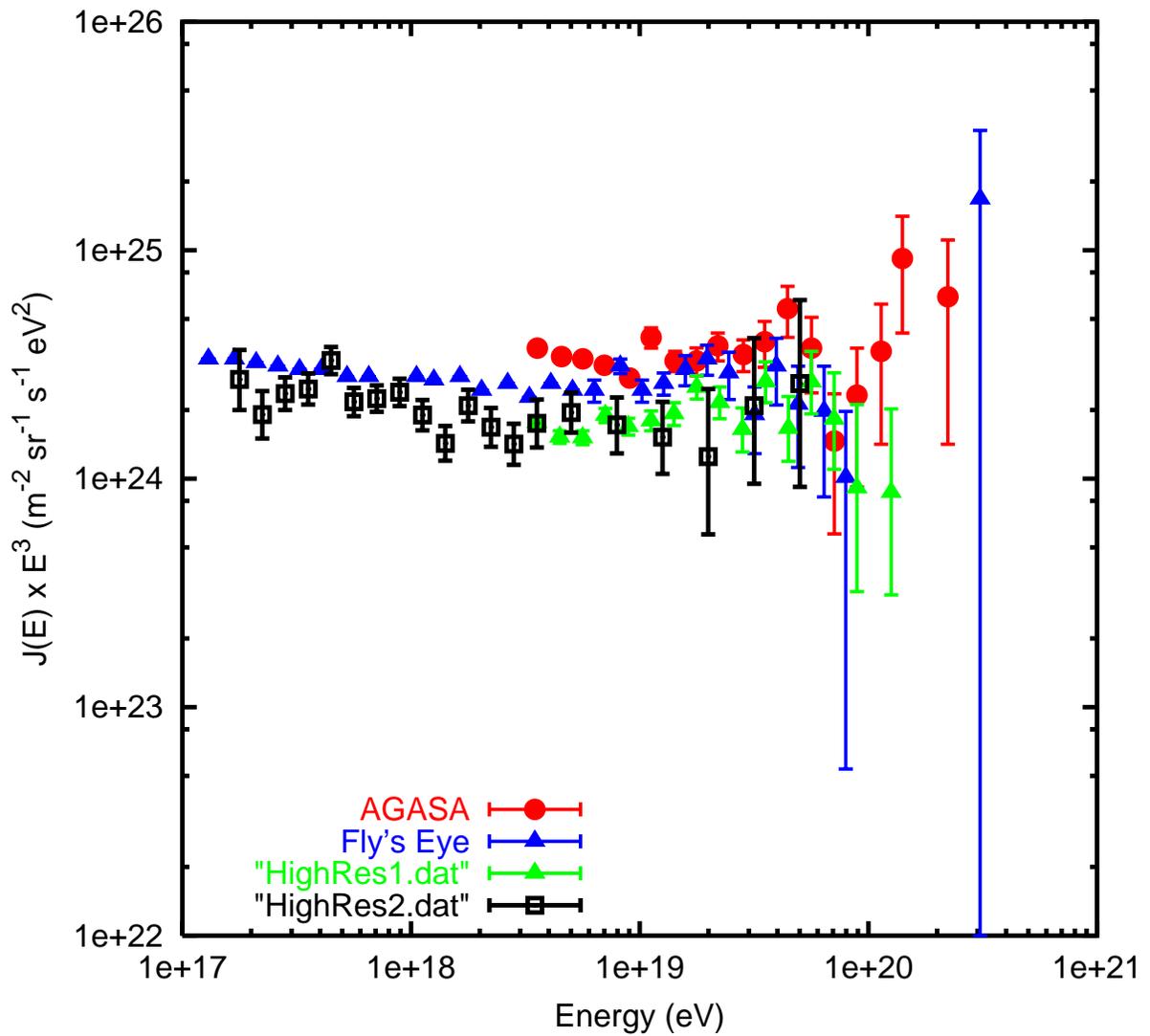}\label{fig3}
\end{center}
\vspace*{-3.0cm} \caption{The experimental data on the UHECRs spectra
from Fly's Eye (blue triangles), AGASA (red circles), HiRes I (green
triangles), and HiRes II (open squares) measurements (from
Ref.~\cite{9}).}
\end{figure}

%\newpage
\begin{figure}[htb]
\begin{center}
\includegraphics[width=\textwidth]{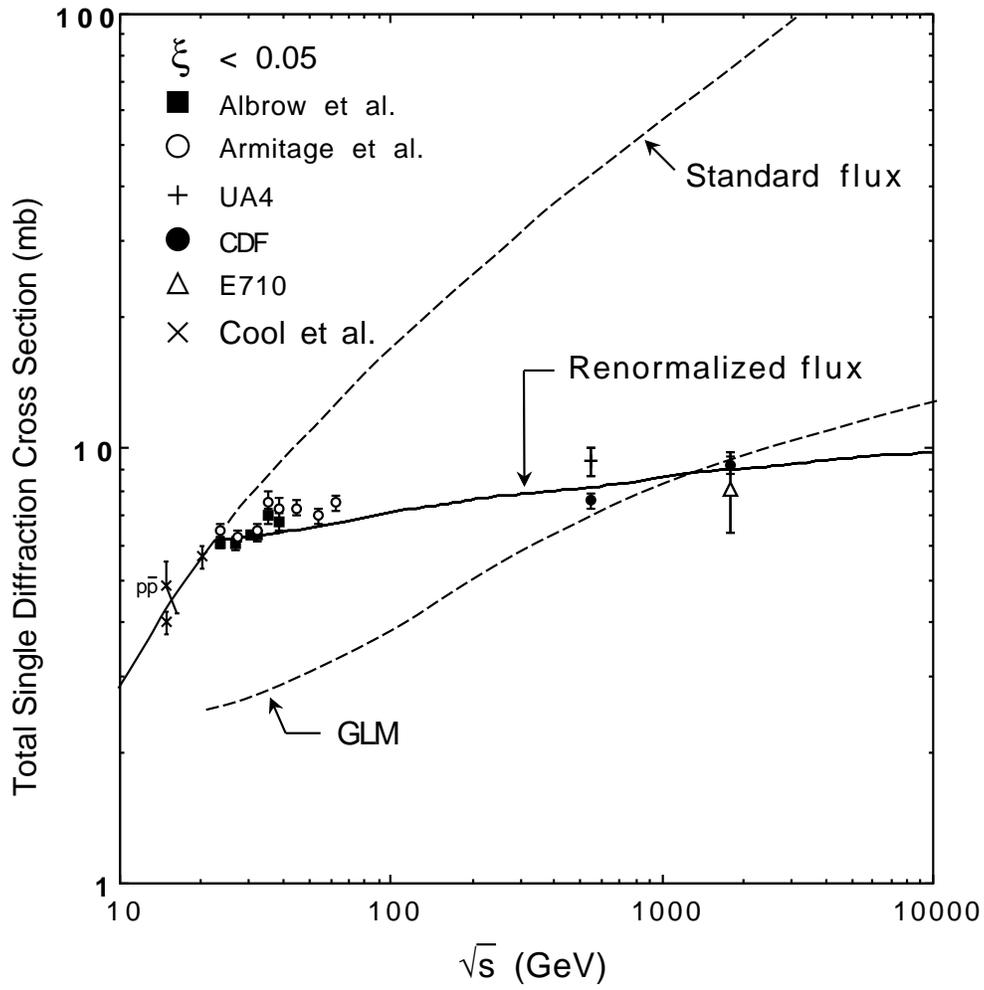}\label{fig4}
\end{center}
\vspace*{-4.0cm} \caption{The total single diffraction cross-sections
for $p(\bar p)+p\rightarrow p(\bar p)+X$ vs $\sqrt{s}$ compared with
the predictions of the renormalized Pomeron flux model of Goulianos
\cite{16} (solid line) and of the model Gostman, Levin and Maor
\cite{17} (dashed line, labelled GLM).}
\end{figure}

%\newpage
\vspace*{3cm}
\begin{figure}[htb]
\begin{center}
\begin{picture}(298,184)
\put(20,10){\includegraphics[]{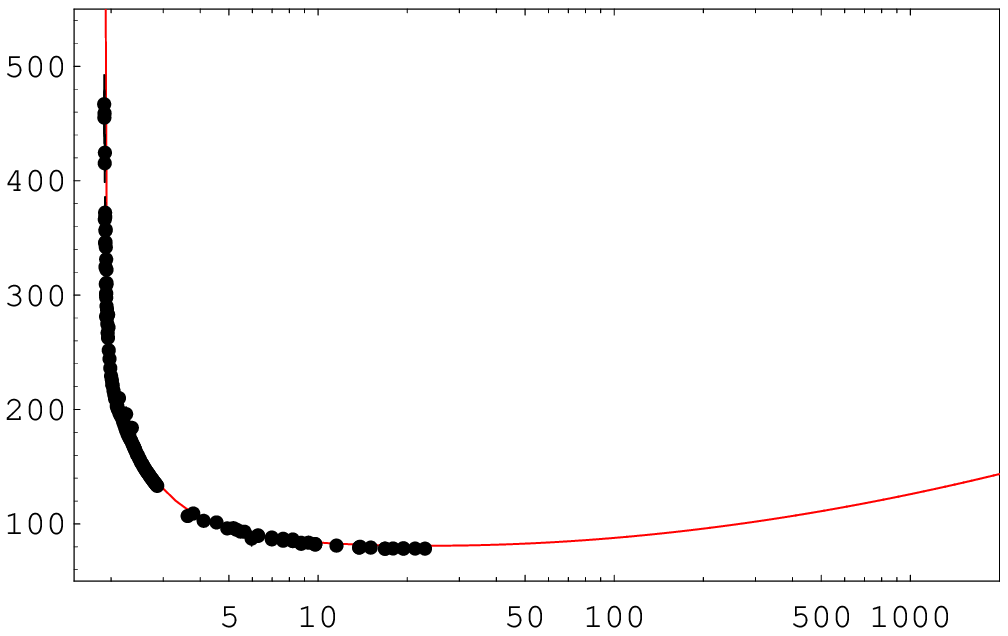}}\label{fig5}
\put(155,0){$\sqrt{s}\, (\rm GeV)$}
\put(0,90){\rotatebox{90}{\large$\sigma^{tot}_{\bar{p}d} (\rm mb)$}}
\end{picture}
\end{center}
\caption{The experimental data on the antiproton-deuteron total
cross-section vs $\sqrt{s}$ \cite{7}. The curve is outcome of
matching the theory and experiment \cite{21}.}
\end{figure}

%\vspace{1cm}

\begin{figure}[htb]
\begin{center}
\begin{picture}(288,184)
\put(20,10){\includegraphics[]{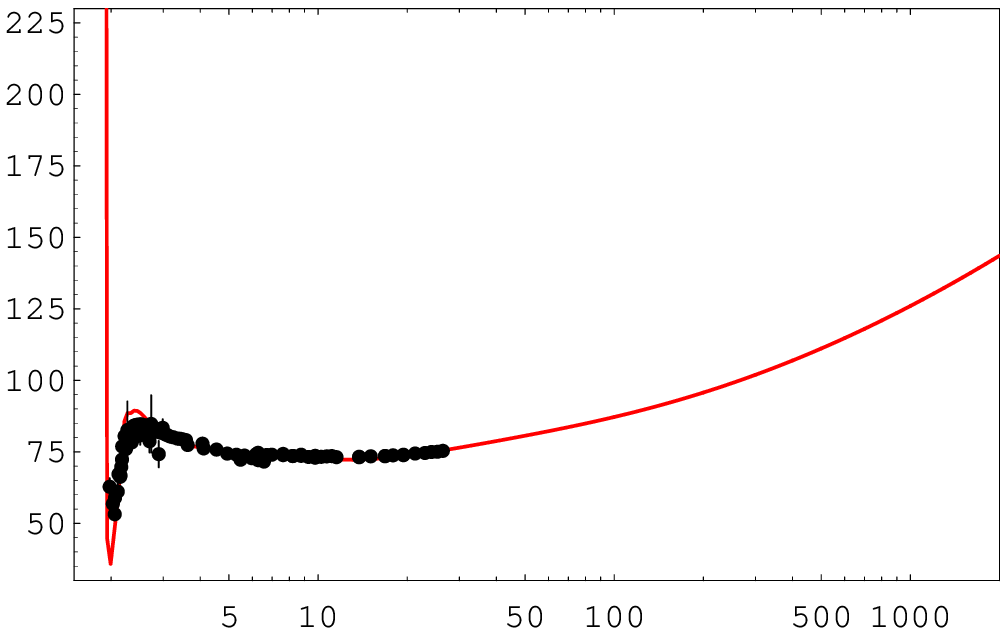}}\label{fig6}
\put(155,0){$\sqrt{s}\,(\rm GeV)$}
\put(0,90){\rotatebox{90}{\large$\sigma^{tot}_{pd} (\rm mb)$}}
\end{picture}
\end{center}
\caption{The experimental data on the proton-deuteron total
cross-section versus $\sqrt{s}$ \cite{7}. The curve is outcome of
matching the theory and experiment \cite{21}.}
\end{figure}

%\newpage
%\vspace*{-1.5cm}
\begin{figure}[htb]
\begin{center}
\begin{picture}(288,204)
\put(15,10){\includegraphics{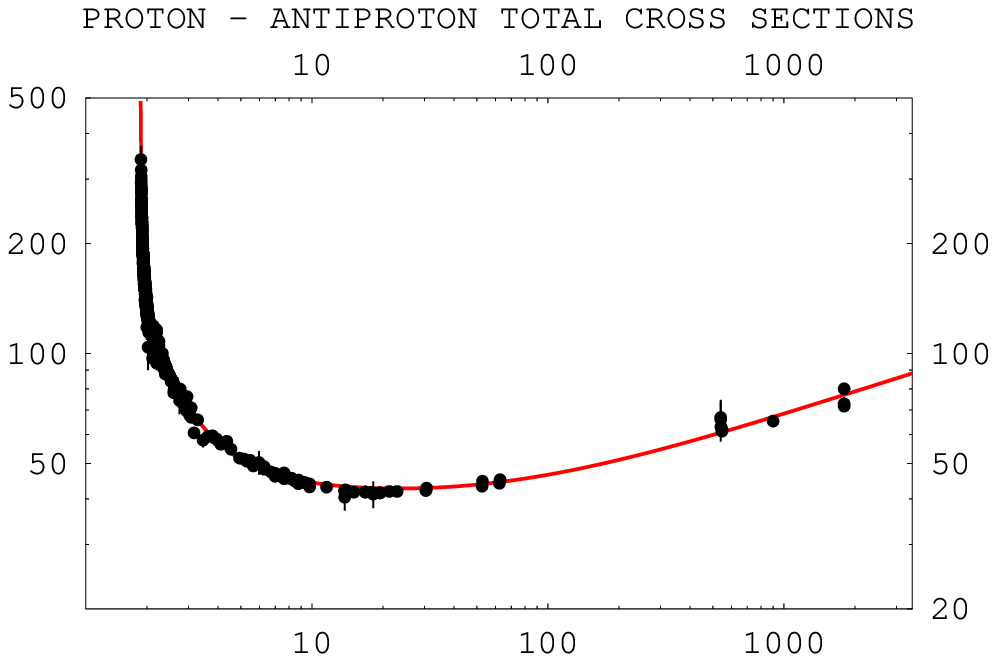}}\label{fig7}
\put(144,0){$\sqrt{s}\,(\rm GeV)$}
\put(-5,87){\rotatebox{90}{\large$\sigma_{p\bar p}^{tot} (\rm mb)$}}
\end{picture}
\end{center}
\caption{The experimental data on the proton-antiproton total
cross-section vs $\sqrt{s}$ \cite{7}. Solid line represents our
global fit \cite {18} to the data. Statistical and systematic errors
added in quadrature.}
\end{figure}

%\vspace{1cm}
\begin{figure}[htb]
\begin{center}
\begin{picture}(288,204)
\put(15,10){\includegraphics{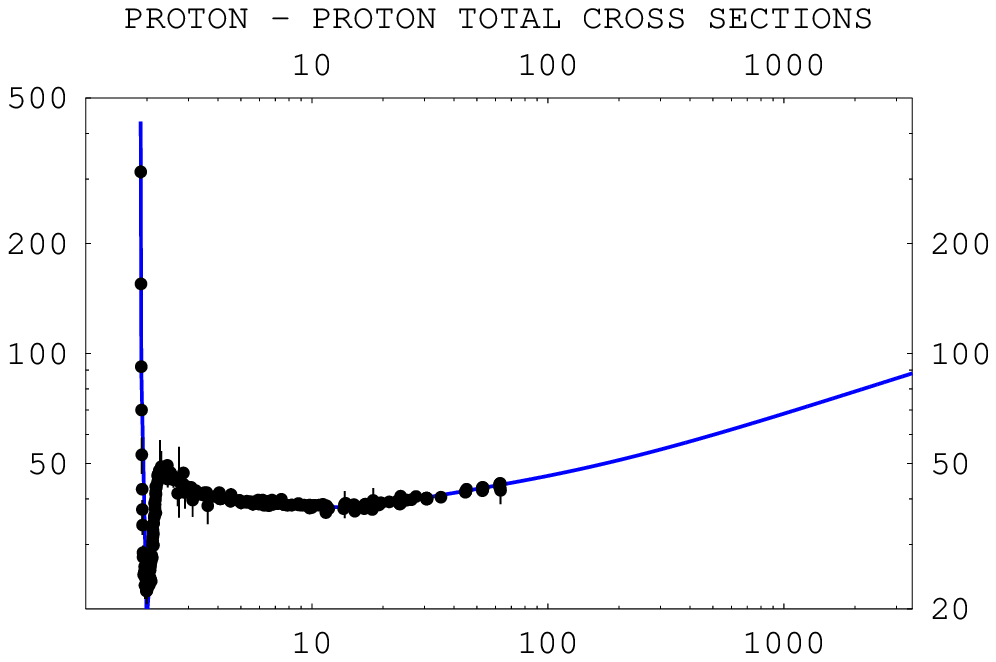}}\label{fig8}
\put(144,0){$\sqrt{s}\, (\rm GeV)$}
\put(-5,87){\rotatebox{90}{\large$\sigma_{pp}^{tot} (\rm mb)$}}
\end{picture}
\end{center}
\caption{The experimental data on the proton-proton total
cross-section versus $\sqrt{s}$ \cite{7}. Solid line represents our
global fit \cite {18} to the data. Statistical and systematic errors
added in quadrature.}
\end{figure}

%\newpage
\begin{figure}
\begin{center}
\begin{picture}(288,188)
\put(15,10){\includegraphics{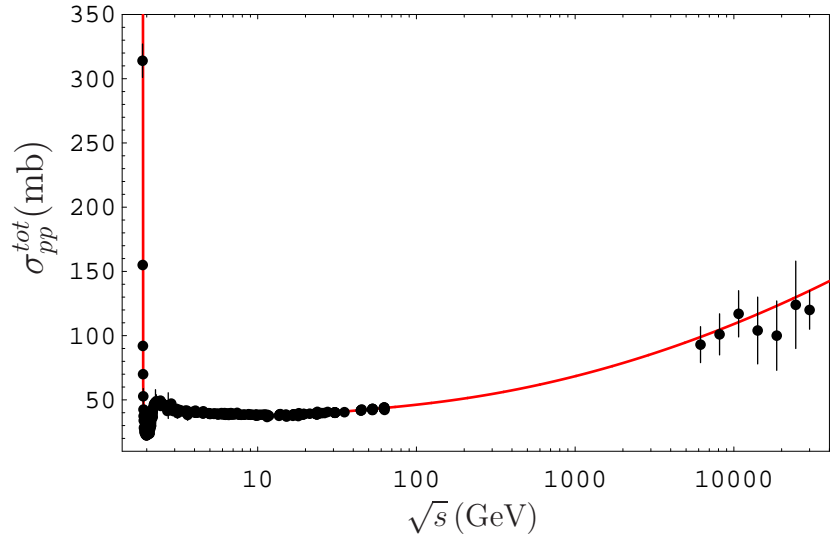}}\label{fig9}
\put(144,0){$\sqrt{s}\, (\rm GeV)$}
\put(-5,95){\rotatebox{90}{\large$\sigma^{tot}_{pp} (\rm mb)$}}
\end{picture}
\end{center}
\caption{The proton-proton total cross-section versus $\sqrt{s}$ with
the cosmic-ray data points from Akeno Observatory and Fly's Eye
Collaboration. Solid line corresponds to our theory predictions
\cite{23}.}
\end{figure}

\vspace{1cm}
\begin{figure}[hbt]
\begin{center}
\begin{picture}(288,188)
\put(15,10){\includegraphics{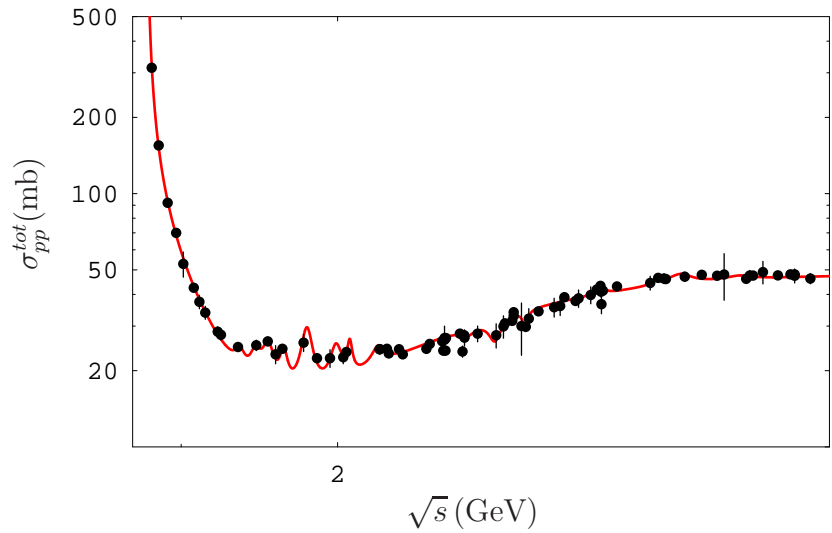}}\label{fig10}
\put(144,0){$\sqrt{s}\,(\rm GeV)$}
\put(-5,95){\rotatebox{90}{$\sigma^{tot}_{pp}(\rm mb)$}}
\end{picture}
\end{center}
\caption{The proton-proton total cross-section versus $\sqrt{s}$ at
low energies. Solid line corresponds to our theory predictions
\cite{24}.}
\end{figure}

%\newpage
\vspace*{1cm}
\begin{figure}[htb]
\begin{center}
\begin{picture}(288,178)
\put(20,10){\includegraphics{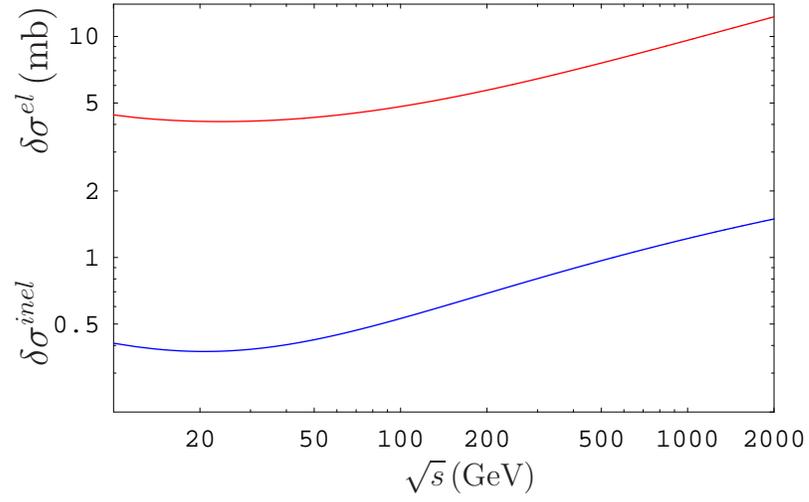}}\label{fig11}
\put(155,0){$\sqrt{s}\, (\rm GeV)$}
\put(8,45){\rotatebox{90}{\large$\delta\sigma^{inel}$}}
\put(8,125){\rotatebox{90}{\large$\delta\sigma^{el} \,(\rm mb)$}}
\end{picture}
\end{center}
\caption{The theoretically calculated elastic and inelastic defects of
total cross section in scattering of (anti)protons from deuterons.}
\end{figure}
\vspace{1cm}
\begin{figure}[htb]
\begin{center}
\begin{picture}(288,172)
\put(20,10){\includegraphics{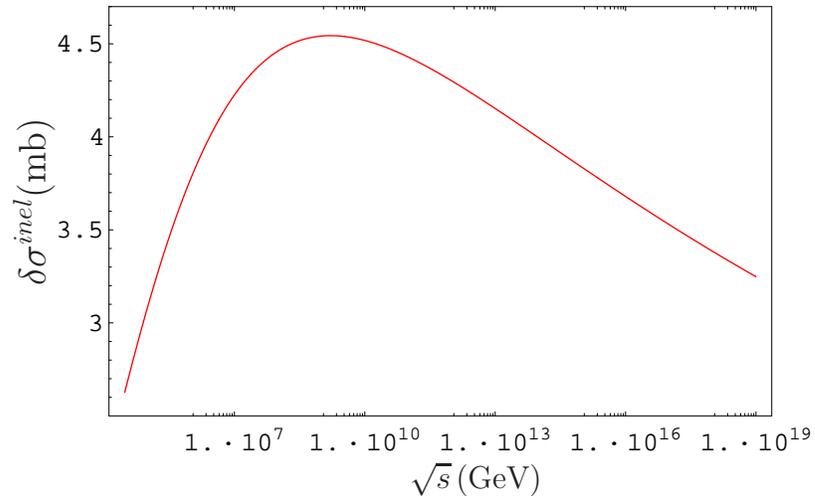}}\label{fig12}
\put(155,0){$\sqrt{s}\,(\rm GeV)$}
\put(5,77){\rotatebox{90}{\large$\delta\sigma^{inel} (\rm mb)$}}
\end{picture}
\end{center}
\caption{The theoretically calculated inelastic defect in the region
of a maximum.}
\end{figure}

%\newpage

\vspace*{1cm}
\begin{figure}[htb]
\begin{center}
\begin{picture}(288,180)
\put(20,10){\includegraphics{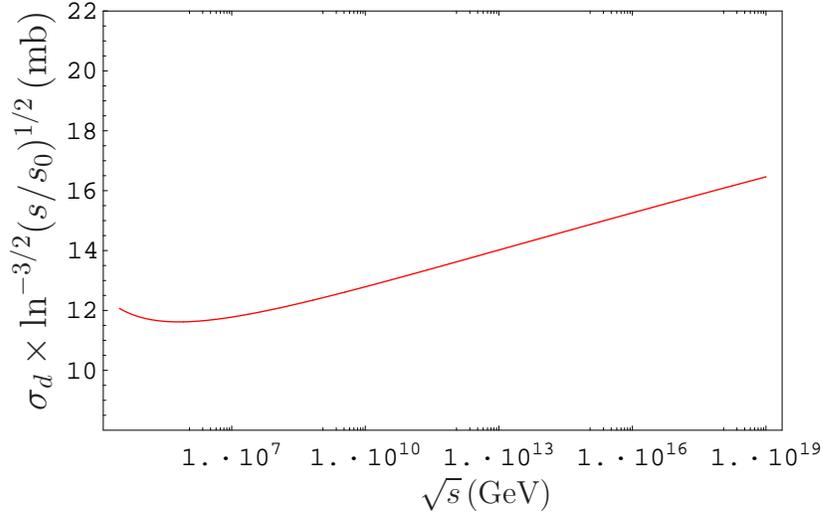}}\label{fig13}
\put(155,0){$\sqrt{s}\, (\rm GeV)$}
\put(0,35){\rotatebox{90}{\large$\sigma_d\times\ln^{-3/2}(s/s_0)^{1/2}
\,(\rm mb)$}}
\end{picture}
\end{center}
\caption{The scaled (anti)proton--deuteron total cross section in the
region of a maximum of the inelastic defect.}
\end{figure}

%\newpage
\begin{figure}[htb]
\begin{center}
\includegraphics[width=0.8\textwidth]{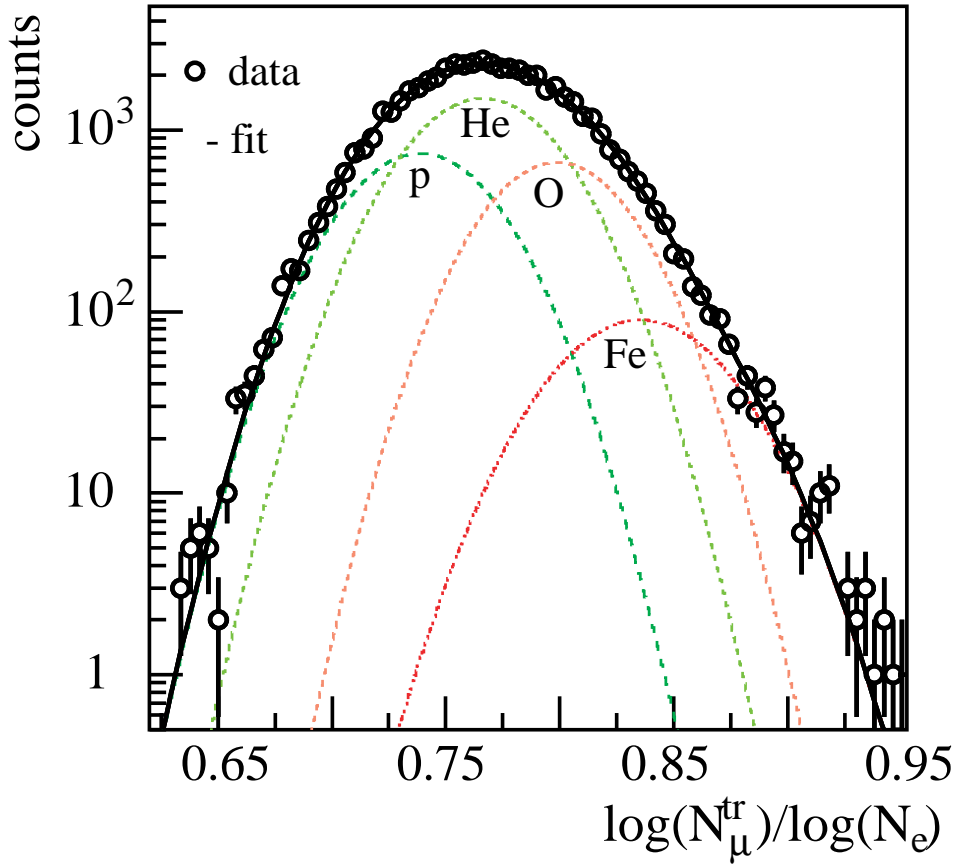}\label{fig14}
\end{center}
\vspace{3cm} \caption{The muon to electron ratio from experimental
data of Kascade compared with simulations for p, He, O and Fe
primaries at energies of about 1 PeV.}
\end{figure}

%\newpage
\begin{figure}[htb]
\begin{center}
\includegraphics[width=\textwidth]{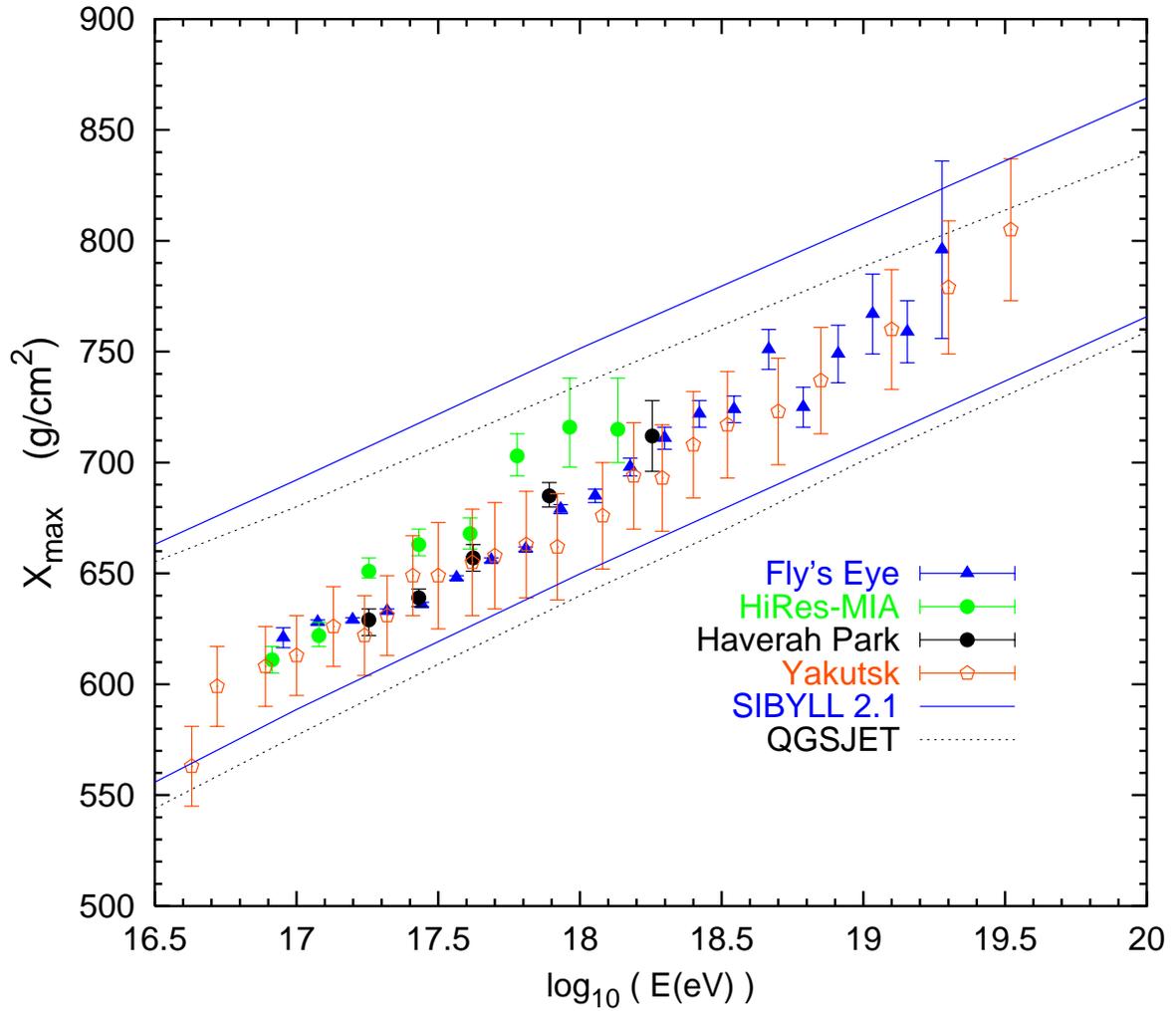}\label{fig15}
\end{center}
\caption{Average depth of shower maximum $<X_{\rm max}>$ vs. energy
compared to the calculated values for protons (upper curves) and iron
primaries (lower curves) in two models (from Ref.~\cite{9}; see
references therein).}
\end{figure}

%\newpage
\begin{figure}[htb]
\begin{center}
\includegraphics[width=\textwidth]{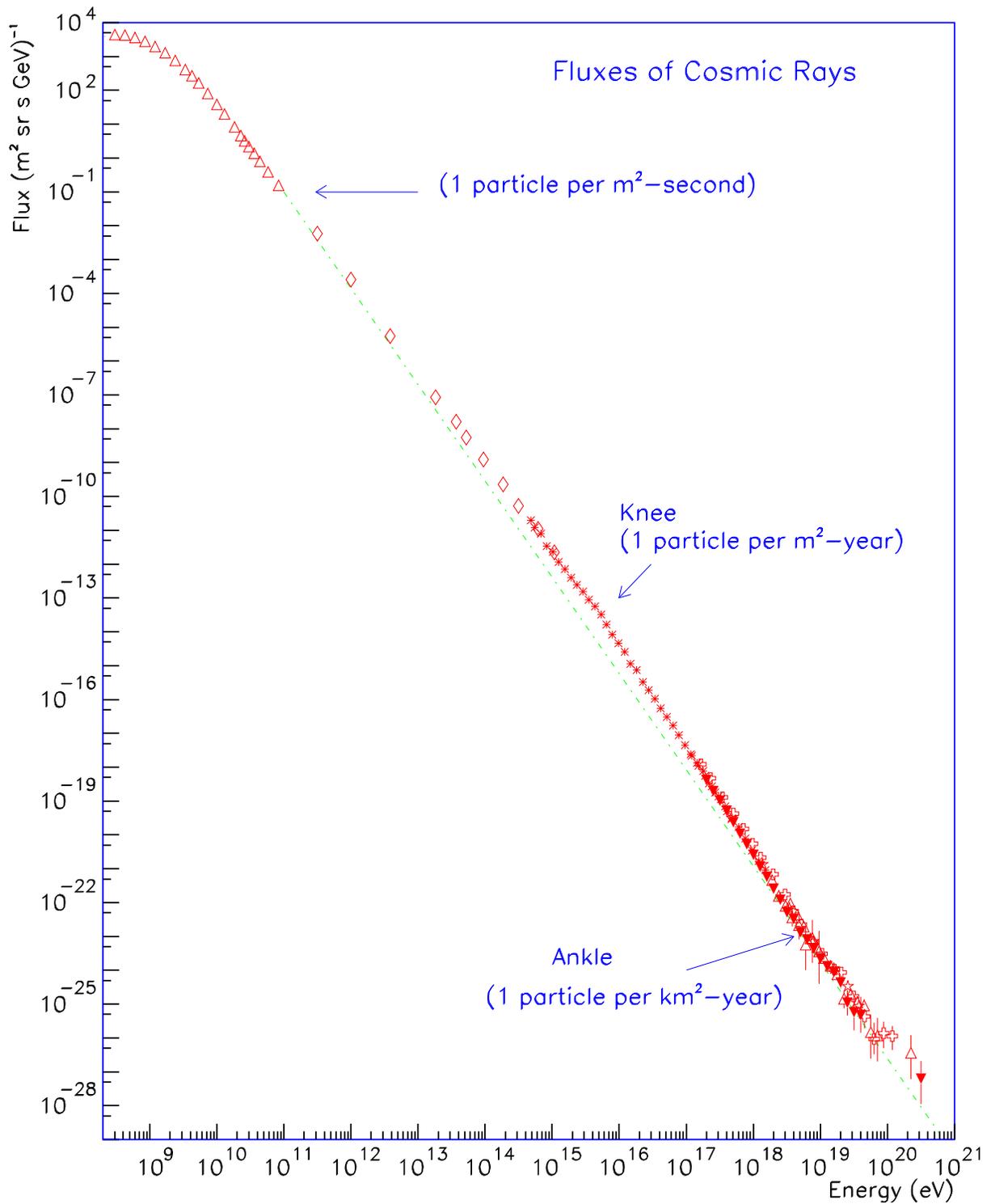}\label{fig16}
\end{center}
\caption{Compilation of measurements of the fluxes of cosmic rays.
The data represent published results of the LEAP, Proton, Akeno,
AGASA, Fly's Eye, Haverah Park, and Yakutsk experiments. The dotted
line shows $E^{-3}$ power-law for comparison. Approximate integral
fluxes (per steradian) are also shown \cite{29}.}
\end{figure}

%\newpage
\begin{figure}[htb]
\begin{center}
\includegraphics[width=\textwidth]{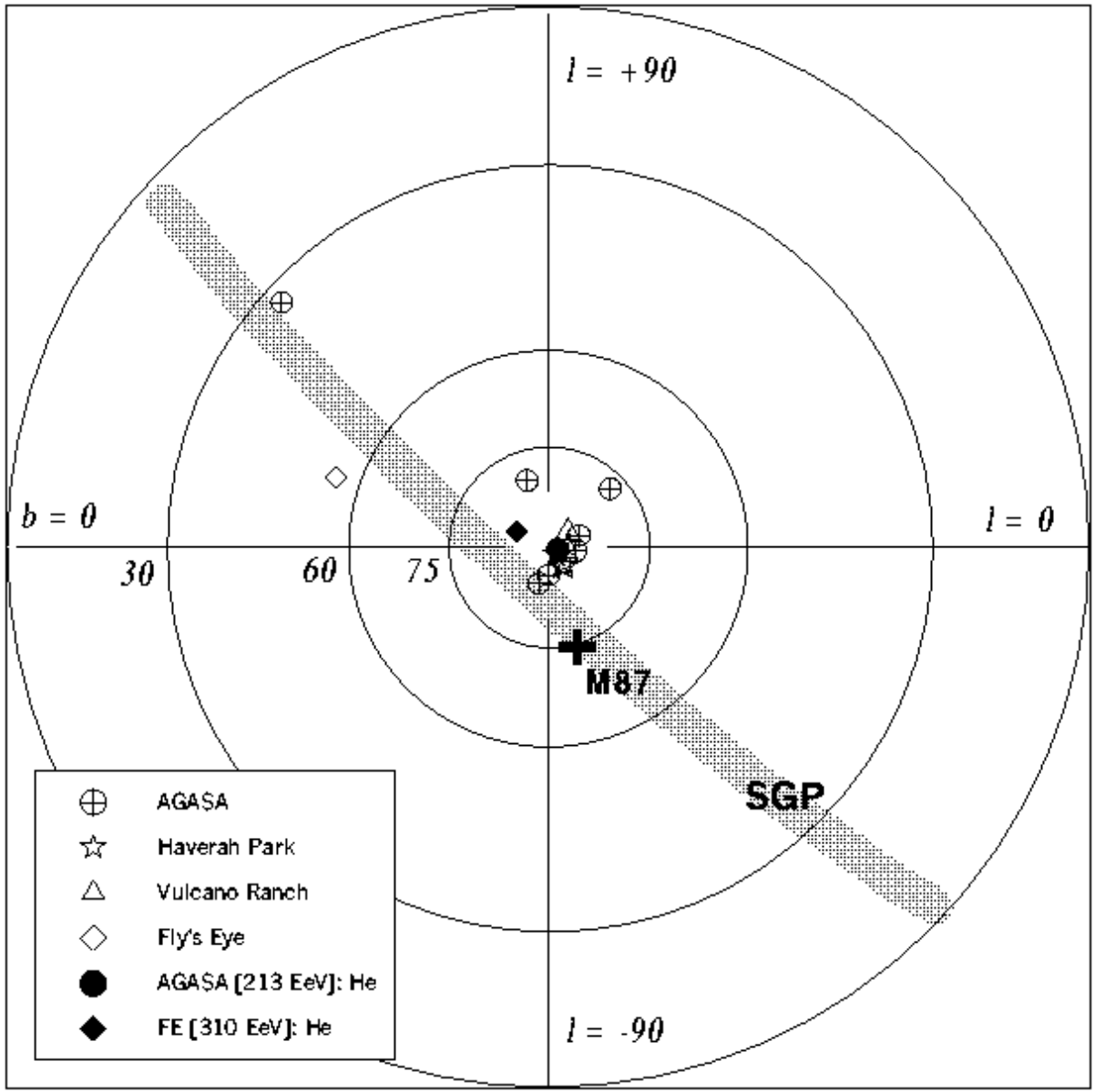}\label{fig17}
\end{center}
\caption{Directions in polar projection of the 13 highest energy
cosmic ray events when they leave the halo of our Galaxy \cite{31}.}
\end{figure}

%\newpage
\begin{figure}[htb]
\begin{center}
\includegraphics[width=\textwidth]{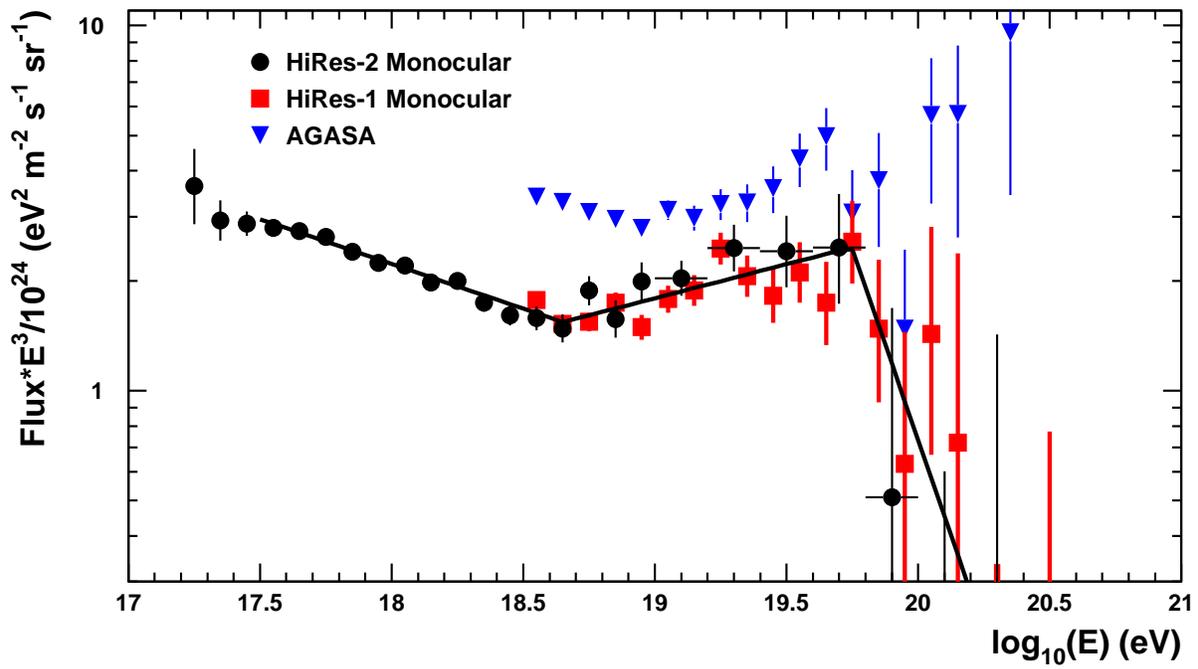}\label{fig18}
\end{center}
\caption{The combined AGASA\&HiRes cosmic ray spectrum presented in
Ref. \cite{32}; see Fig. 3 there.}
\end{figure}

\end{document}